\documentclass[preprint,12pt]{elsarticle}
\usepackage{amsmath,amssymb,amsthm,amsfonts,appendix,hyperref,subcaption}

\biboptions{sort&compress}

\newcommand{\abs}[1]{\left\lvert #1 \right\rvert}
\newcommand{\ceil}[1]{\left\lceil #1 \right\rceil}
\newcommand{\hfrac}[2]{#1/#2}
\def\O{\mathcal{O}}
\def\N{\mathbb{N}}
\def\R{\mathbb{R}}
\def\Tr{\text{Tr}}
\def\Det{\text{Det}}
\newtheorem{prop}{Proposition}
\newtheorem{theorem}{Theorem}
\newtheorem{corollary}{Corollary}
\newtheorem{lemma}{Lemma}

\def\mycopyrightnotice{%
  {\footnotesize {\textcopyright \enspace 2025. This manuscript version is made available under the CC-BY-NC-ND 4.0 license \href{https://creativecommons.org/licenses/by-nc-nd/4.0/}{https://creativecommons.org/licenses/by-nc-nd/4.0/}}
  \gdef\mycopyrightnotice{}
}}

\journal{Mathematical Biosciences}

\begin{document}

\begin{frontmatter}

\title{The Mean-Field Survival Model for Stripe Formation in Zebrafish Exhibits Turing Instability\tnoteref{1}}
\author[Boston University]{Robert Jencks\corref{CorAuth}}
\affiliation[Boston University]{organization={Department of Mathematics \& Statistics, Boston University},
            addressline={665 Commonweath Ave}, 
            city={Boston},
            postcode={02215}, 
            state={MA},
            country={United States of America}}
\cortext[CorAuth]{Corresponding author.}
\ead{jencksr@bu.edu}
\tnotetext[1]{\mycopyrightnotice}

\begin{abstract}
Zebrafish have been used as a model organism in many areas of biology, including the study of pattern formation. The mean-field survival model is a coupled ODE system describing the expected evolution of chromatophores coordinating to form stripes in zebrafish. This paper presents analysis of the model focusing on parameters for the number of cells, length of distant-neighbor interactions, and rates related to birth and death of chromatophores. We derive the conditions on these parameters for a Turing bifurcation to occur and show that the model predicts patterns qualitatively similar to those in nature. 
    
In addition to answering questions about this particular model, this paper also serves as a case study for Turing analysis on coupled ODE systems. The qualitative behavior of such coupled ODE models may deviate significantly from continuum limit models. The ability to analyze such systems directly avoids this concern and allows for a more accurate description of the behavior at physically relevant scales.
\end{abstract}

\begin{highlights}
\item Characterization of Turing instability in a ring-based ODE model for zebrafish.
\item Quantitative analysis for 4 main parameters, including any finite number of cells, \(N\).
\item Stability curves for large \(N\) differ substantially from previous PDE model.
\item Critical melanophore projection properties similar to those observed \textit{in vivo}.
\item Case study for Turing analysis on ring-based ODE models without taking PDE limit.
\end{highlights}


\end{frontmatter}

\section{Introduction}

The mechanisms by which patterns form is an active area of study in fields ranging from theoretical to lab sciences. The mathematical study of morphogensis was transformed by the discovery of the Turing mechanism in 1952. Turing presented the linear stability analysis (LSA) of various reaction-diffusion models, demonstrating that a difference in diffusion rates can lead to spatially periodic patterns \cite{Turing}. These models included systems with two or three chemical components, continuous and discrete spatial domains, and analysis on rings and spheres. Much of the mathematical interest in the subject has been on the development of theory for the continuum case \cite{Eckhaus, Edelstein-Keshet, Murray1993, Walgraef, Epstein, Kondo2010, Vansaarloos, Prigogine,Krause}. For these conventional reaction-diffusion systems, short-range activation and long-range inhibition is a common mechanism for pattern formation \cite{Turing,Krause}. This terminology refers to the idea that there must be an agent which diffuses slowly and promotes activity and the other agent diffuses quickly and inhibits activity. In models replacing diffusion with a non-local operator, it is possible for pattern formation to occur even when the activator has longer characteristic interaction length than the inhibitor \cite{Kondo2017}. Models utilizing such non-local operators can display significant sensitivity to spatial dimension, potentially allowing pattern formation in some dimensions while disallowing pattern formation in others \cite{Jewell}. 

In discrete models where diffusion is replaced by graph Laplacian coupling terms, similar behavior can occur \cite{Turing,Ide,Wolfrum,nakao,McCullen}. In the case of two components on a ring, the only patterns that occur are stationary, spatially periodic states \cite{Turing}. The simple nature of this network allows for the explicit characterization of the onset of instability and the description of a chemical wavelength which provides estimates for the observed wavelength in actual steady states \cite{Turing}. In contrast, the study of large, complex networks has shown the existence of multi-stability of stripes and localized patterns \cite{Ide,Wolfrum,nakao,McCullen}. These network models differ from the ring model in \cite{Turing} by studying typical behavior in random graphs (usually Erd\H{o}s-R\'{e}yni graphs) \cite{Ide,Wolfrum,nakao,McCullen}, rather than on a ring of fixed size. These results are powerful in their ability to describe networks which are too complex and unstructured for explicit analysis. For such networks, the behavior of one specific graph is less informative than the average behavior of many graphs. Overall, Turing-like instabilities occur in a diverse set of conditions across a large variety of models making them a cornerstone of pattern formation analysis.

This flexibility is particularly applicable to the problem of stripe formation in zebrafish. These stripes are formed by the distribution of pigmented cells, called chromatophores. In zebrafish, there are three primary chromataphores: iridophores (silver), xanthophores (yellow), and melanophores (black). Of these, iridophores have much greater mobility enabling them to ``diffuse" more significantly than xanthophore or melanophores \cite{Singh, Mahalwar, Parichy, Frohnhofer, Takahashi}. However, melanophores and xanthophores have been observed to interact with each other in a manner that influences stripe regrowth \cite{Nakamasu}. Melanophores are capable of extending projections from their cell body which reach multiple cell diameters away. These projections tend to favor movement towards xanthophores and are typically at least three body lengths long \cite{Hamada}. The projections appear to engage in a signaling interaction, leading to a difference in cell birth and death rates after ablation and by extension influence the formation of stripes \cite{Nakamasu}. Some observations have indicated that iridophores may not be necessary for stripe formation \cite{Hirata} while others have indicated that they are essential \cite{Frohnhofer}. These observations have lead to a variety of modeling approaches primarily interested in pattern formation \cite{Bullara, Konow, Volkening2015}. Some models include neither cell migration nor iridiphore influence, instead focusing on xanthophore-melanophore interactions in lattice-based environments \cite{Bullara, Konow}. Other models use an agent based approach to incorporate the movement of cells, but not the influence of iridophores \cite{Volkening2015}. Such agent based models have been developed further still to include iridophores and to differentiate the behavior of loosely- and densely-packed cells \cite{Volkening2018}. Readers interested in further review of the state of research into stripe formation in zebrafish specifically will appreciate Kondo, Watanabe, and Miyazawa's 2021 review article on the subject \cite{Kondo2021}.

\begin{figure}[htbp]
    \centering
    \includegraphics[scale=0.75]{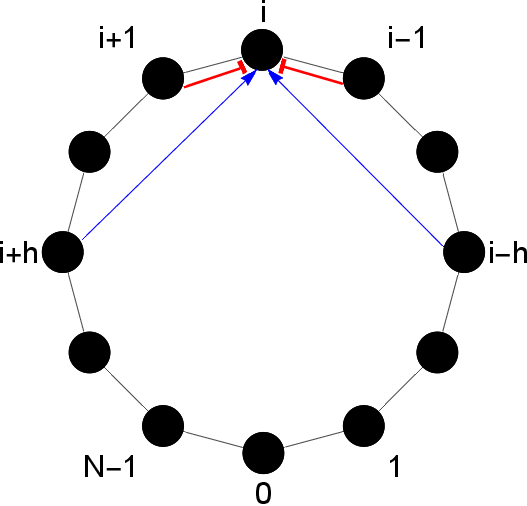}
    \caption{Schematic of ring geometry of model \eqref{1D ODE}. Red arrows represent short-range interactions and blue represent long-range interactions.}
    \label{fig: Interaction Schematic}
\end{figure}

In this paper, we study the following model for the interaction of melanophores and xanthophores, modified from the models presented in \cite{Konow}:
\begin{subequations}
\label{1D ODE}
    \begin{align}
        X_{j}'=&1-X_{j}-M_{j}-X_{j}\left(\tfrac{M_{j-1}+M_{j+1}}{2}\right);\label{1D X ODE}\\
        M_{j}'=&(b+1)(1-X_{j})-(b+d+1)M_{j}-M_{j}\left(\tfrac{X_{j-1}+X_{j+1}}{2}\right)\nonumber\\
        &+dM_{j}\left(\tfrac{X_{j-h}+X_{j+h}}{2}\right).\label{1D M ODE}
    \end{align}
\end{subequations}
The system consists of \(2N\) ODEs for variables \(X_{j}\) and \(M_{j}\) with \(0\leq j \leq N-1\). Cells are modelled on a ring, so indices are modulo \(N\). \(X_{j}\) and \(M_{j}\) represent the probability of finding a xanthophore or melanophore at location \(j\) respectively. In equation \eqref{1D X ODE}, the first three terms combined correspond to the probability of location \(j\) being empty and a xanthophore appearing with relative rate \(1\). The remaining, non-linear term in \eqref{1D X ODE} represents the death of xanthophores due to the presence of nearby melanophores at relative rate \(1\). The first two terms in equation \eqref{1D M ODE} represents a combination of multiple effects; growth from site \(j\) being empty contributes \((b+1)(1-X_{j}-M_{j})\) and natural death contributes \(-dM_{j}\). The first of the non-linear terms in \eqref{1D M ODE} models death from nearby xanthophores while the second models promotion of melanophore growth from xanthophores at distance \(h\) away. Short-range interactions, \(j\pm 1\), are inhibitory while the long-range interactions, \(j\pm h\), are excitatory on melanophores when interacting with xanthophores. Figure \ref{fig: Interaction Schematic} contains an example schematic of the  overall model structure. The remaining parameters, \(b\) and \(d\), control the birth rate and death rate of the melanophores respectively. In addition to modeling stripe formation in zebrafish, the model is representative of a variety of cell-compartment models. The analysis of this model is a detailed case study for the analysis of this broad class of models.

\subsection{Brief Discussion of Previous Survival Model Analysis}

Konow et al. presented many results on the numeric and analytic study of three models for stripe formation in zebrafish \cite{Konow}. These three models are a continuous-time, discrete-space stochastic model (the survival model), an ODE model (the mean-field survival model) derived from the survival model, and a PDE model (the continuous mean-field survival model) derived from the mean-field survival model. The original survival model is motivated by experimental observations \cite{Hamada,Nakamasu,Mahalwar} and the modeling approach of the promotion model \cite{Bullara}. The decision to consider cells at fixed locations and to omit the influence of iridophores reflects the conditions in zebrafish fins, which display patterns despite a lack of iridophores and restricted cell movement \cite{Konow,Singh}. The full survival model has three types of stochastic state transitions resulting from short-range interactions, long-range interactions, and spontaneous changes without external interactions. Spontaneous changes are used to model the birth of melanophore and xanthophores as well as the deaths of xanthophores. Melanophore deaths are not included in these interactions because of the prominent influence xanthophores have on melanophore death rates \cite{Nakamasu}. The short range interactions describe additional deaths for melanophores and xanthophores due to overcrowding. The long-range interactions describe the different death rates of melanophores depending on the presence of xanthophores at distant locations to model the signals melanophores appear to receive from their projections \cite{Hamada}. Xanthophore projections are significantly shorter and less common than malanophore projections, so only melanophore projection effects are included in the model \cite{Hamada}.

The mean-field survival model represents the master equation of the survival model under the assumption that there is no correlation across space of the stochastic transitions. Model \eqref{1D ODE} is derived by a simple change in notation and some simplifying assumptions from the mean-field survival model presented below \cite{Konow}.
\begin{subequations}
\label{Konow ODEs}
    \begin{align}
        \tfrac{d\langle X_{j}\rangle}{dt}=&b_{X}(1-\langle X_{j}\rangle-\langle M_{j}\rangle)-d_{x}\langle X_{j}\rangle-\tfrac{s_{M}}{2}\langle X_{j}\rangle\left(\langle M_{j-1}\rangle+\langle M_{j+1}\rangle\right);\label{Konow X ODE}\\
        \tfrac{d\langle M_{j}\rangle}{dt}=&b_{M}(1-\langle X_{j}\rangle-\langle M_{j}\rangle)-\tfrac{s_{X}}{2}\langle M_{j}\rangle\left(\langle X_{j-1}\rangle+\langle X_{j+1}\rangle\right)\nonumber\\
        &-\tfrac{d_{MX}}{2}\langle M_{j}\rangle\left(\langle X_{j-h}\rangle+\langle X_{j+h}\rangle\right)\label{Konow M ODE}\\
        &-\tfrac{d_{M}}{2}\langle M_{j}\rangle\left(\langle S_{j-h}\rangle+\langle S_{j+h}\rangle+\langle M_{j-h}\rangle+\langle M_{j+h}\rangle\right).\nonumber
    \end{align}
\end{subequations}
Here, \(\langle X_{j}\rangle\) and \(\langle M_{j} \rangle\) refer to the expected values of Boolean random variables indicating the presence or absence of xanthophores and melanophores at a given cell location. The quantity \(S_{i}\) is the probability that location \(i\) is empty, meaning \(S_{i}=1-X_{i}-M_{i}\). \(b_{X}\) and \(b_{M}\) represent birth rates, \(d_{x}\) and \(d_{M}\) natural death rates, and \(s_{x}\) and \(s_{M}\) short-range interaction death rates of xanthophores and melanophores respectively. \(d_{MX}\) represents a different death rate from melanophores with a long-range interaction with a xanthophore.

The primary analysis of the model presented in Konow et al. used the simplifying assumptions \(s_{x}=s_{M}=1\), \(b_{X}=1\), \(d_{X}=0\), and \(d_{MX}=0\) \cite{Konow}, and these simplifying assumptions are reflected in system \eqref{1D ODE}. These assumptions allow for direct comparison with the promotion model which made similar assumptions and are maintained in the current work. There is a biological motivation to some of these assumptions. For example, melanophore stripes which have adjacent xanthophore stripes ablated have been observed to die more quickly than the case where adjacent xanthophores are present \cite{Nakamasu}. The difference was so large that in the former case, the melanophore density decreased by \(\approx 10\%\) in the same period that melanophore density increased \(\approx 10\%\) in the latter case \cite{Nakamasu}. This indicates that \(d_{MX}<<d_{M}\) and motivates the assumption that \(d_{MX}=0\). The other assumptions, like setting xanthophore death rate to \(0\), are made to make the analysis tractable. We also make a change in notation to simplify many formulas in our results; we define \(b=b_{M}-1\), \(d=d_{M}\), \(X_{j}(t)=\langle X_{j}(t)\rangle\), and \(M_{j}(t)=\langle M_{j}(t)\rangle\). Rewriting system \eqref{Konow ODEs} using the assumptions and change of notation yields system \eqref{1D ODE}.

 Here, we briefly describe how this paper extends their analysis further. In their analysis, Konow et al. studied the survival model and the mean-field survival model numerically, while reserving the analytic work for the continuous mean-field survival model \cite{Konow}. They performed a LSA of the continuous model and characterized the conditions for a Turing bifurcation. This was compared with the onset of pattern formation in the numerical simulations of a variety of scenarios. We build upon this by performing the LSA on the mean-field survival model directly. The pattern formation analysis is more difficult than the continuous mean-field PDE, but the behavior is less abstracted from the original model. This provides new information about the original discrete models as well as capturing phenomena which do not appear in the LSA of the continuous mean-field survival model.

This model differs from the network models studied in \cite{Turing,Ide,Wolfrum,nakao,McCullen} primarily in the way that cells are coupled to each other. In all of those models, the node-to-node coupling is linear \cite{Turing,Ide,Wolfrum,nakao,McCullen}, while the coupling terms in the present model appear only in the non-linearity. Additionally, the coupling is given by an average of selected cells, which is 
distinct from the graph Laplacian used in the other models \cite{Turing,Ide,Wolfrum,nakao,McCullen}. Finally, the non-local/distant neighbor coupling in the melanophore equation is different from the nearest neighbor coupling of Turing \cite{Turing}. The model may be interpreted as an abstract network with only nearest neighbor interactions on a new graph in addition to the interpretation as a ring with a non-local interaction. This description is more similar to the complex networks in \cite{Ide,Wolfrum,nakao,McCullen}, but differs in the explicit and highly symmetric structure of the resulting network.

System \eqref{1D ODE} admits two homogenous equilibria,
\begin{align}
    \left(X,M\right)=\left(1,0\right),\left(\tfrac{d}{b+d},\tfrac{b}{b+2d}\right).\label{Hom Eq}
\end{align}
As \(X\) and \(M\) are probabilities, we restrict the parameter regime to \(b> 0\) and \(d> 0\) so that these values both remain between \(0\) and \(1\). While the original model admits all \(b\geq -1\), only the \((1,0)\) equilibrium is physically relevant for \(-1\leq b <0\). The focus of this paper is on the coexistence equilibrium, which is only physically relevant for \(b\geq 0\). Additionally, as the cell locations are on a ring, we restrict the discrete variables to \(1\leq h\leq \hfrac{N}{2}\). Values of \(h\) larger than \(\hfrac{N}{2}\) correspond to projections wrapping around the ring and interacting with the same locations as in the \(N-h\) case.

\subsection{Overview of Objective and Results}

The goals of this paper are to characterize the pattern formation properties of model \eqref{1D ODE} and to demonstrate the benefits of performing LSA on the system of coupled ODEs directly. The results presented in this paper focus on the LSA of the non-trivial homogeneous equilibria from \eqref{Hom Eq} and the spatially periodic patterns that emerge from the Turing instabilities. In particular, we study the bifurcations with respect to four main parameters of \eqref{1D ODE}, including all physically relevant values of the kinetics parameters \(b\) and \(d\), the cell count, \(N\), and the non-local interaction distance, \(h\). All four parameters have significant influences on the stability of the base state \eqref{Hom Eq} and formation of periodic solutions. We analytically determine the dependence of the neutral stability curve on each parameter and identify their individual effects. We show that the model predicts stripes to form for a large range of kinetic parameter values so long as the total cell count and non-local interaction distance are sufficiently large. This indicates that the length of melanophore projections and the overall size of the zebrafish must be above some minimum threshold for stripes to develop. In fact, the minimum interaction distance predicted by this model is exactly the minimum length of melanophore projections typically observed in zebrafish. The analysis also provides bounds on the relationship between projection length, cell count and stripe width. These bounds predict that, over a large parameter regime, the width of stripes relative to projection length is similar in the model and in nature. Similarly, the observed width of stripes and number of stripes provide estimates for the \(h\) and \(N\) values where the model is most similar to reality.

Additionally, we study interesting limiting cases and their effect on the bifurcation analysis. The limit \(N\to\infty\) provides a valuable point of comparison between the ODE dynamics and the dynamics of the PDE derived in Konow et al for the same limit. The discrepancy between the PDE behavior and \(N\to \infty \) ODE behavior demonstrates potential pitfalls in the usage of continuum limits to study even large coupled ODE systems. For appropriately chosen \(d\) and \(h\), the homogeneous state in the PDE system is unstable on an unbounded interval of \(b\) values, \(b_{0}<b\). However, the \(N\to\infty\) limit of the corresponding intervals in the ODE system is always bounded, \(b_{0}<b<b_{1}\), a dramatic difference in linear stability behavior. The asymptotes for either \(d\) large or \(b\) and \(d\) both large provide a simple approximation for the bifurcation curve. As \(h\) increases, empirical evidence suggests that the bifurcation curve is increasingly well-approximated by its asymptotes; visually distinguishing the curves on \((b,d)\in[0,8]\times[0,16]\) is already difficult for \(h=11\) (see Figure \ref{fig: Bifurcation no asymptotes}).

This paper also serves as a case study for the analysis of pattern formation in systems of ODEs in a manner similar to Turing's original work. This is valuable because often we are interested in models with finite values of \(N\) which may significantly deviate from the dynamics of large \(N\) PDE limits. While much closer to the finite \(N\) behavior, the limiting ODE behavior still fails to capture some detail about the precise onset of instability, even when considering \(N=100\). The main results in this paper do require that \(N\) be larger than some critical value, but this is not generally so large as to be unrealistic. In fact, model \eqref{1D ODE} demonstrates pattern formation for parameters as small as \(N=6\) and \(h=3\). So, any zebrafish which has developed enough to be studied is well past the minimum size for this model to allow stripes to form. This type of minimal parameter bound can only be found by analyzing all values of \(N\) and \(h\) on the ODE system directly and would be lost in a continuum limit approach.

The results indicate that many of the zebrafish stripe formation trends observed in nature can be explained by signaling between melanophores and distant neighbors. In particular, the width of stripes relative to projection length and minimum typical projection length in the model are similar to those observed in nature. More generally, this suggests that distant neighbor signaling is a viable mechanism for pattern formation. It also indicates that lattice-based models are capable of accurately describing pattern formation in scenarios where pigment cell movement is believed to be non-critical in the formation of patterns.

\section{Results}
\subsection{Main Results}

Let \(\left(X_{\text{H}}(t),M_{\text{H}}(t)\right)\) be any homogeneous solution to system \eqref{1D ODE} and define \(u_{j}(t)=X_{j}(t)-X_{\text{H}}(t)\) and \(v_{j}(t)=M_{j}(t)-M_{\text{H}}(t)\). Denote the discrete Fourier transform of \(u_{j}(t)\) and \(v_{j}(t)\) by \(\hat{u}_{k}(t)\) and \(\hat{v}_{k}(t)\) respectively. Then \(\left\{X_{j}(t),M_{j}(t)\right\}_{i=0}^{N-1}\) solves system \eqref{1D ODE} if and only if \(\left\{\hat{u}_{k}(t),\hat{v}_{k}(t)\right\}_{k=0}^{N-1}\) solves system \eqref{ODE DFT}
\begin{align}
    \begin{bmatrix}
        \hat{u}_{k}'\\
        \hat{v}_{k}'
    \end{bmatrix}=&\begin{bmatrix}
        -M_{\text{H}}(t)-1&-x_{k}X_{\text{H}}(t)-1\\
        \left(dT_{h}(x_{k})-x_{k}\right)M_{\text{H}}(t)-b-1&(d-1)X_{\text{H}}(t)-b-d-1
    \end{bmatrix}\begin{bmatrix}
        \hat{u}_{k}\\
        \hat{v}_{k}
    \end{bmatrix}\label{ODE DFT}\nonumber\\
    &+\begin{bmatrix}
        -\big(\hat{u}\ast x_{k}\hat{v}\big)_{k}\\
        -\big(x_{k}\hat{u}\ast\hat{v}\big)_{k}+d\big(T_{h}(x_{k})\hat{u}\ast\hat{v}\big)_{k}
    \end{bmatrix},
\end{align}
where \(\ast\) denotes circular convolution on \(N\) elements, \(x_{k}=\cos\left(2\pi \hfrac{k}{N}\right)\), and \(T_{h}(x)\) is the \(h\)-order Chebyshev polynomial of the first kind. We will refer to the matrix in equation \eqref{ODE DFT} as \(\hat{L}_{k}\) and take care that \((X_{\text{H}}(t),M_{\text{H}}(t))\) are clear from context.

For the trivial equilibrium \(\left(1,0\right)\), LSA shows that 
\begin{align*}
    \Tr\left(\hat{L}_{k}\right)=&-b-3,\\
    \Det\left(\hat{L}_{k}\right)=&1-(b+1)x_{k}.
\end{align*}
For all admissible parameters, \(\Tr\left(\hat{L}_{k}\right)<0\). The linear stability is therefore governed entirely by \(\Det\left(\hat{L}_{k}\right)\). From this, we conclude the following proposition:

\begin{prop}\label{prop: Equilibrium1}
    The equilibrium \(\left(1,0\right)\) is linearly unstable to spatially periodic perturbations in the subspace of discrete Fourier mode \(k\) for each \(k\) satisfying
    \begin{align*}
        x_{k}\geq \tfrac{1}{b+1},
    \end{align*}
    and linearly stable to all other perturbations.
\end{prop}

The pure xanthophore equilibrium is predicted to be physically unobservable in the presence of melanophores as the homogeneous mode, \(k=0\), is always unstable.

For the nontrivial equilibrium \(\left(\tfrac{d}{b+d},\tfrac{b}{b+2d}\right)\), the matrix \(\hat{L}_{k}\) becomes
\begin{align*}
    \hat{L}_{k}=&\begin{bmatrix}
        \frac{-2(b+d)}{b+2d} & \frac{-\left(d(x_{k}+1)+b\right)}{b+d}\\
        \frac{b\left(dT_{h}(x_{k})-x_{k}\right)}{b+2d}-(b+1) & \frac{-(b+1)(b+2d)}{b+d}
    \end{bmatrix}.
\end{align*}
If \(\lambda\) is an eigenvalue of this matrix, then \(\Vec{\xi}\), defined below, is a corresponding eigenvector.
\begin{align*}
    \Vec{\xi}=&\begin{bmatrix}
        (b+2d)\left(d(x_{k}+1)+b\right)\\
        -\left(2(b+d)^{2}+\lambda\right)
    \end{bmatrix}.
\end{align*}
For the values of \(b\) and \(d\) where this equilibrium is physically relevant, the first component is strictly positive, but the second component may be positive or negative depending on the value of \(\lambda\). From this observation, we conclude the following proposition:

\begin{prop}\label{prop: Unstable Eigenvector}
    If the equilibrium \(\left(\tfrac{d}{b+d},\tfrac{b}{b+2d}\right)\) has any real, non-negative eigenvalues, then the corresponding eigenvectors must have components with opposite signs.
\end{prop}

The significance of this proposition, is that it implies that the melanophore and xanthophore stripes are out of phase with each other. This means that, if stripes do form, they will always alternate between high melanophore density and high xanthophore density, as is observed in nature.

To determine whether or not there are such real, non-negative eigenvalues, we note that 
\begin{align*}
    \Tr\left(\hat{L}_{k}\right)=&\frac{-2(b+d)^{2}-(b+1)(b+2d)^{2}}{(b+d)(b+2d)},\\
    \Det\left(\hat{L}_{k}\right)=&\frac{F\left(x_{k},T_{h}(x_{k}),b,d\right)}{(b+d)(b+2d)},
\end{align*}
where \(F:\R^{4}\to\R\) is defined as follows:
\begin{align}\label{F def}
    F(x,y,b,d)=&-bdx^{2}+\left(bd^{2}y-b(b+d)-d(b+1)(b+2d)\right)x\\
    &+bd(b+d)y+(b+1)(b+d)(b+2d).\nonumber
\end{align}
An alternative characterization of \(F\) which will be useful in later analysis is
\begin{align}\label{F Alt}
    F(x,y,b,d)=&b\left(b^{2}+bd(-x+y+3)+d^{2}\left((x+1)y-2(x-1)\right)\right)\\
    &-(x-1)\left(b^{2}+(x+3)bd+2d^{2}\right).\nonumber
\end{align}

The equilibrium undergoes a Turing bifurcation when \(F=0\) with \(F>0\) corresponding to linear stability and \(F<0\) indicating an instability to appropriate spatially periodic perturbations. As such, it is important to understand the characteristics of the \(F=0\) level set in different planes. One particularly important parameter regime for this characterization is the set \(\mathcal{P}\) defined by
\begin{align}\label{P Def}
    \mathcal{P}=\left\{(b,d)\in\R^{2}:0< d,0< b <d\right\}.
\end{align}
This region will be crucial in the statement of Theorems \ref{thm: Discrete Bifurcation} and \ref{thm: Continuous Bifurcation} classifying the onset of Turing instability with respect to the discrete and continuous parameters respectively. Biologically, this region is approximately the region where the melanophore death rate is faster than the spontaneous birth rate. The content of Theorems \ref{thm: Discrete Bifurcation} and \ref{thm: Continuous Bifurcation} suggests that this is a requirement for pattern formation to occur. It is likely that there exists a more complex set of physiological conditions which are necessary and that this simplified set of conditions holds only for the simplifying assumptions made in this analysis. It is important to note that, while these results require \(N\) be sufficiently large, this does not mean that the value of \(N\) must be exceptionally large. After the presentation of the theorems, we also explore the case of the minimal value of \(N\).

\begin{figure}[htbp]
\begin{subfigure}{0.49\textwidth}
    \centering
    \includegraphics[width=0.9\linewidth]{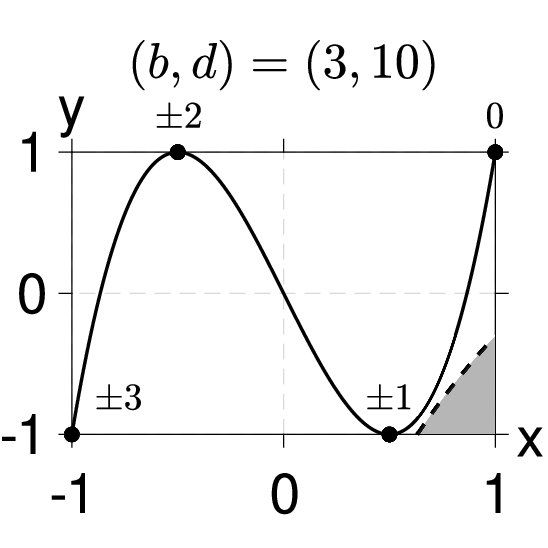}
    \subcaption{}
    \label{fig: h3N6b3d10 Cheb plot}
\end{subfigure}
\begin{subfigure}{0.49\textwidth}
    \centering
    \includegraphics[width=0.9\linewidth]{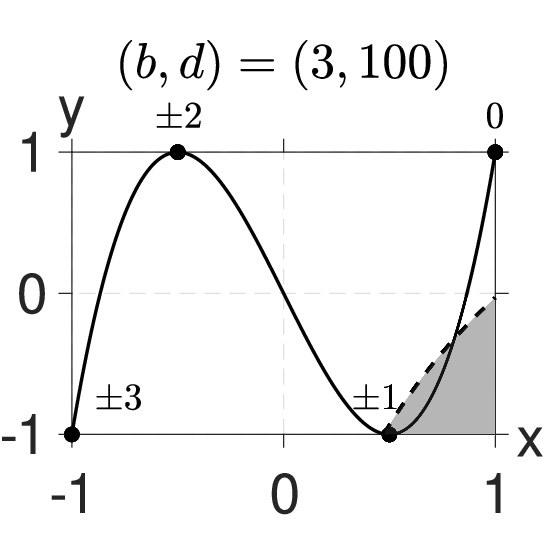}
    \subcaption{}
    \label{fig: h3N6b3d100 Cheb plot}
\end{subfigure}
\caption{The curves \(y=T_{h}(x)\) (solid black) and \(F(x,y,b,d)=0\) (dashed black) in the \(xy\)-plane. The set with \(F<0\) is the shaded region below and to the right of the \(F=0\) curve. Turing instability occurs when one or more of the points \((x_{k},T_{h}(x_{k}))\) (black markers) lie in the region with \(F<0\) for some \((b,d)\). All markers are labeled with the mode(s) that they correspond to. For both plots, \(h=3\) and \(N=6\). \textbf{(\subref{fig: h3N6b3d10 Cheb plot})} For \((b,d)=(3,10)\), the equilibrium is linearly stable. \textbf{(\subref{fig: h3N6b3d100 Cheb plot})} For \((b,d)=(3,100)\), the \(k=\pm1\) modes are unstable. Changing the value of \(b\) has a similar effect of changing the shape of the \(F=0\) curve.}
\label{fig: h3N6 Cheb plots}
\end{figure}

\begin{figure}[htbp]
\begin{subfigure}[t]{0.49\textwidth}
    \centering
    \includegraphics[width=0.9\linewidth]{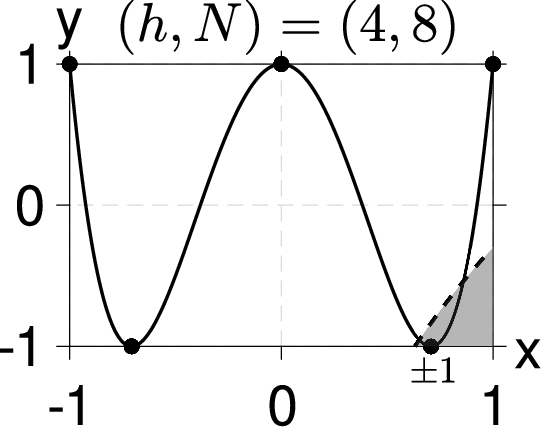}
    \subcaption{}
    \label{fig: h4N8b3d10 Cheb plot}
\end{subfigure}
\begin{subfigure}[t]{0.49\textwidth}
    \centering
    \includegraphics[width=0.9\linewidth]{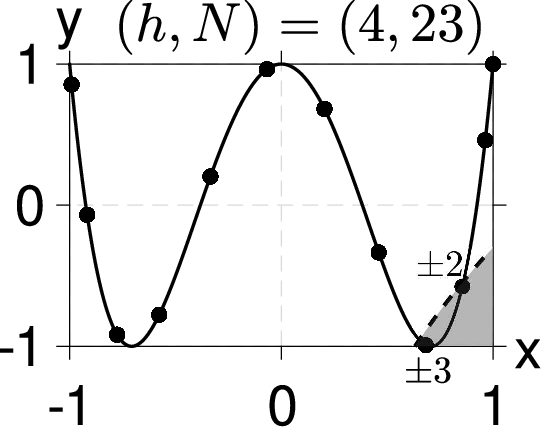}
    \subcaption{}
    \label{fig: h4N23b3d10 Cheb plot}
\end{subfigure}

\begin{subfigure}[t]{0.98\textwidth}
    \centering
    \includegraphics[width=0.9\linewidth]{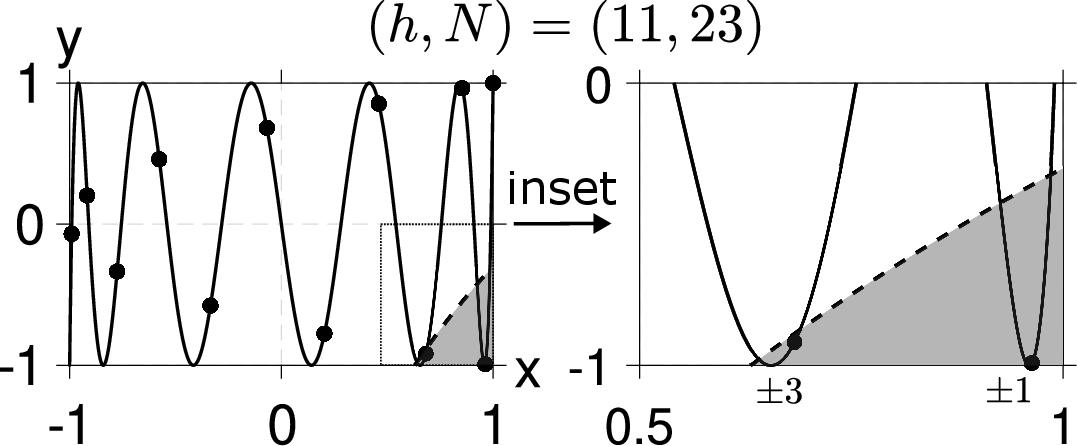}
    \subcaption{}
    \label{fig: h11N23b3d10 Cheb plot}
\end{subfigure}
\caption{The curves \(y=T_{h}(x)\) (solid black) and \(F(x,y,b,d)=0\) (dashed black) in the \(xy\)-plane. The set with \(F<0\) is the shaded region below and to the right of the \(F=0\) curve. Turing instability occurs when one or more of the points \((x_{k},T_{h}(x_{k}))\) (black markers) lie in the region with \(F<0\) for some \((b,d)\). The rightmost marker corresponds to the \(k=0\) mode, the next marker corresponds to the \(k=\pm 1\) modes, etc. The modes which are unstable are labeled in all plots (in the inset for Figure \ref{fig: h11N23b3d10 Cheb plot}). \((b,d)=(3,10)\) for all plots. \textbf{(\subref{fig: h4N8b3d10 Cheb plot})} For \(h=4\) and \(N=8\), the \(k=\pm 1\) modes are unstable. \textbf{(\subref{fig: h4N23b3d10 Cheb plot})} For \(h=4\) and \(N=23\), the \(k=\pm 2,\pm 3\) modes are unstable. Different choices of \((b,d)\) may cause only the \(k=\pm 2\) or only the \(k=\pm 3\) modes to be unstable. \textbf{(\subref{fig: h11N23b3d10 Cheb plot})} For \(h=11\) and \(N=23\), the \(k=\pm 1,\pm 3\) modes are unstable. Unlike the unstable modes in \textbf{(\subref{fig: h4N23b3d10 Cheb plot})}, \(k=\pm 3\) modes being unstable implies the \(k=\pm 1\) modes are unstable. Inset depicts the region \([\hfrac{1}{2},1]\times[-1,0]\) marked by a dotted rectangle in the original image.}
\label{fig: b3d10 Cheb plots}
\end{figure}

In Figures \ref{fig: h3N6 Cheb plots} and \ref{fig: b3d10 Cheb plots}, we can develop an intuition for the style of argument employed in the proofs of the main theorems. These theorems establish some necessary and some sufficient conditions on the parameters for spatially periodic instability to occur as well as bounds for the Fourier modes which have the strongest linear instability. In the \(xy\)-plane, the key objects are the curve \(y=T_{h}(x)\), the points \((x,y)=(x_{k},T_{h}(x_{k}))\) embedded in this curve, and the region where \(F<0\). There is instability if and only if \(F(x_{k},T_{h}(x_{k}),b,d)<0\) for some \(k\), so we seek to show instability by showing one of the points \((x,y)=(x_{k},T_{h}(x_{k}))\) lies in the region where \(F<0\). To establish the desired bounds, we must bound the region on which the minimum of \(F(x_{k},T_{h}(x_{k}),b,d)\) occurs.

Figure \ref{fig: h3N6 Cheb plots} explores the effect in the \(xy\)-plane of changing \(d\) while \(h\) and \(N\) are held constant. Generally speaking, the further the point \((b,d)\) is from the boundary of \(\mathcal{P}\), the larger the region \(F<0\) becomes. If this region expands to include one of the points \((x,y)=(x_{k},T_{h}(x_{k}))\), then the equilibrium will become unstable to perturbations with the corresponding Fourier mode. This motivates the first major step in the proofs of the main theorems: establishing bounds on the \(F=0\) level set in the \(xy\)-plane.

Figure \ref{fig: b3d10 Cheb plots} demonstrates the effect in the \(xy\)-plane of changing \(h\) and \(N\) while \(b\) and \(d\) are held constant. An increase in \(h\) leads to more local extrema in the curve \(y=T_{h}(x)\). This forces the rightmost local minimum closer to \(x=1\), potentially causing the curve \(y=T_{h}(x)\) to enter the region where \(F<0\). Increasing \(N\) increases the density of the points \((x,y)=(x_{k},T_{h}(x_{k}))\) in the curve \(y=T_{h}(x)\). These observations give the exact tools needed for the other major steps of the proofs of instability: choose \(h\) large enough for the curve \(y=T_{h}(x)\) to enter the region where \(F<0\) and choose \(N\) large enough that some point \((x,y)=(x_{k},T_{h}(x_{k}))\) lies on that section of the curve. Figure \ref{fig: h11N23b3d10 Cheb plot} also provides intuition for the establishment of bounds on the modes with the largest linear instability. The negative level sets of \(F\) have qualitatively similar shape to the \(F=0\) level set presented, indicating that points lower and further right in the \(xy\)-plane will be more linearly unstable. This can be used to show the minimum value of \(F\) on the curve \(y=T_{h}(x)\) must lie between the rightmost local minimum and the rightmost root, leading to bounds on the modes with the largest linear instability.

We are now ready to present and discuss the main theorems of the paper.

\begin{theorem}\label{thm: Discrete Bifurcation}

    If \((b,d)\in\mathcal{P}\), then \(\exists H_{1}\geq 3\) depending only on \((b,d)\) and \(N_{1}\in \N\) depending on \((b,d)\) and \(h\) such that \(\forall h\geq H_{1}\) and \( N\geq N_{1}\), the equilibrium \(\left(\tfrac{d}{b+d},\tfrac{b}{b+2d}\right)\) is unstable to appropriate spatially periodic perturbations in the subspace of discrete Fourier mode(s) \(k\) for some \(k\neq 0\).
    
    Additionally, \(\exists N_{2}\in\N\) depending on \((b,d)\) and \(h\) such that \(\forall N\geq N_{2}\), the value(s) of \(k\)  corresponding to the largest positive eigenvalue satisfies
    \begin{align}
    \label{Bounds1}
        \ceil{\tfrac{N}{4h}}\leq k \leq \ceil{\tfrac{N}{2h}},
    \end{align}
    where \(\ceil{\cdot}\) denotes the ceiling function. 
    
    If \((b,d)\notin\mathcal{P}\), then the equilibrium \(\left(\tfrac{d}{b+d},\tfrac{b}{b+2d}\right)\) is linearly stable.
\end{theorem}

In summary, for each \((b,d)\in\mathcal{P}\), choosing \(h\) and \(N\) large enough produces instability to spatially periodic perturbations. There are also well-defined bounds on the wavelengths of these perturbations based on the ratio of long-range interaction distance to the total number of cell locations. For \((b,d)\notin \mathcal{P}\), there is no such instability. Biologically, being in \(\mathcal{P}\) means that, in the absence of distant xanthophores to stabilize a melanophore stripe, the melanophores will die faster than they regrow and the stripe with begin to fade. This is similar to the behavior observed in the ablation experiments in \cite{Nakamasu}. If the death rate is higher than the birth rate, the model predicts that beyond some critical projection length and fish size, stripes will always eventually form. 

The bounds on \(k\) indirectly provide bounds on the wavelength of stripes, \(\hfrac{N}{k}\), measured in cell-diameters. The model predicts that stripe width should always be between \(2\) and \(4\) times the typical projections length. This agrees with observations by Hamada et al that projections can be nearly half of the stripe width \cite{Hamada}. The width of stripes has been observed to be \(7\)-\(12\) times that of cells \cite{Nakamasu,Volkening2015}. This suggests a physical upper bound of \(h\leq 6\). The projections are typically three times the width of the cell body \cite{Hamada}, indicating a physical lower bound of \(3\leq h\). This should be compared to the range of \(h\) values for which the model produces stripes with the correct width relative to cell diameter. The width of the stripes being \(7\)-\(12\) times that of a cell means that \(N\) should be approximately \(14\)-\(24\) times \(k\) for the model to produce the desired stripes. Inequality \eqref{Bounds1} indicates that this occurs for \(\hfrac{7}{2}\leq h \leq 13\). The range of realistic projection lengths for the model contains the range of lengths typically observed in nature, indicating it is capable of producing realistic stripes.

The previous theorem gives a characterization of the behavior that is expected when \((b,d)\) has already been fixed and \(h\) is free to vary. It is also important to consider the behavior when \(h\) is fixed and \((b,d)\) is free to vary.

\begin{theorem}\label{thm: Continuous Bifurcation}

    \(\forall h\geq 3\), \(\exists N_{3}\in \N\) depending only on \(h\) such that \(\forall N\geq N_{3}\), \(\exists k\neq 0\) such that the equilibrium \(\left(\tfrac{d}{b+d},\tfrac{b}{b+2d}\right)\) is unstable to appropriate spatially periodic perturbations in the subspace of discrete Fourier mode(s) \(k\) and some \((b,d)\). The set of \((b,d)\) for which the \(k\) mode is unstable satisfies
    \begin{align*}
        \mathcal{U}_{k}(h,N)=\left\{(b,d)\in\R_{+}^{2}:F\left(x_{k},T_{h}(x_{k}),b,d\right)<0\right\}\subset\mathcal{P}.
    \end{align*}
    For \(h=1,2\), the equilibrium is linearly stable for all admissible values of \((b,d)\).
    
    Additionally, \(\exists N_{4}\in\N\) depending only on \(h\) such that \(\forall N\geq N_{4}\), the value(s) of \(k\) which first loses stability and/or corresponding to the largest positive eigenvalue satisfies
    \begin{align}
    \label{Bounds2}
        \ceil{\tfrac{N}{4h}}\leq k \leq \ceil{\tfrac{N}{2h}}.
    \end{align}
    The value of \(k\) which first loses stability and/or corresponds to the largest positive eigenvalue may depend on \((b,d)\).
\end{theorem}

\begin{figure}[htbp]
\begin{subfigure}[t]{0.49\textwidth}
    \centering
    \includegraphics[width=0.9\linewidth]{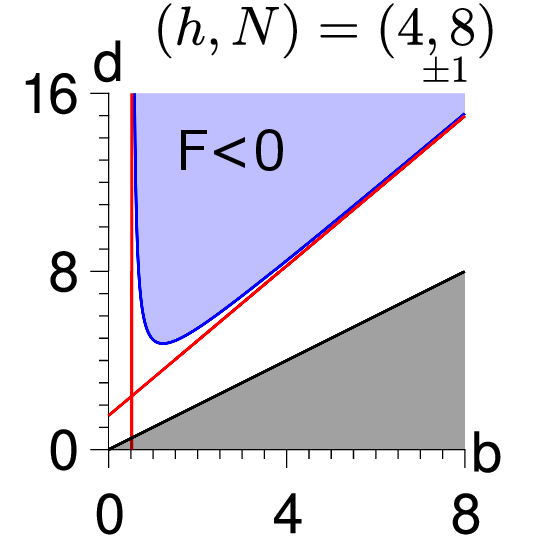}
    \subcaption{}
    \label{fig: h4N8 Bifurcation}
\end{subfigure}
\begin{subfigure}[t]{0.49\textwidth}
    \centering
    \includegraphics[width=0.9\linewidth]{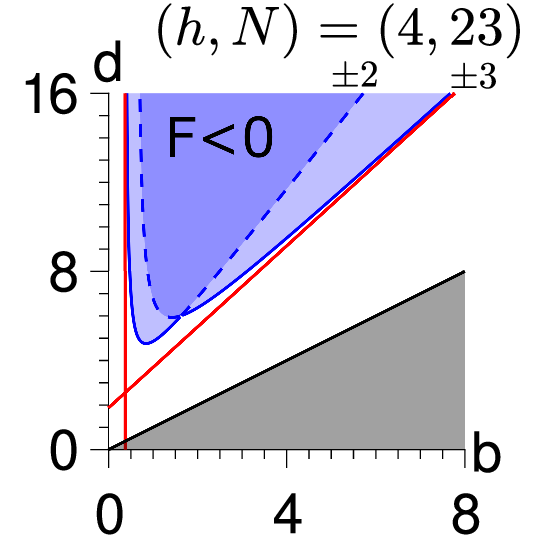}
    \subcaption{}
    \label{fig: h4N23 Bifurcation}
\end{subfigure}

\begin{subfigure}[t]{0.98\textwidth}
    \centering
    \includegraphics[width=0.45\linewidth]{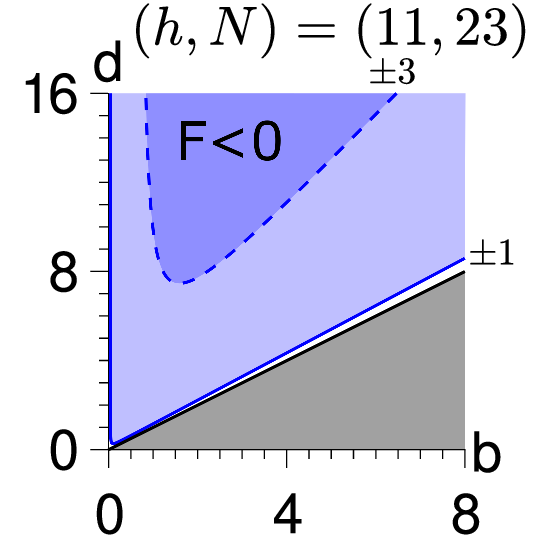}
    \subcaption{}
    \label{fig: h11N23 Bifurcation}
\end{subfigure}
\caption{The curves, \(F(x_{k},T_{h}(x_{k}),b,d)=0\), in the \(bd\)-plane. The bifurcation curve (solid blue) represents the boundary where the equilibrium is neutrally stable. The neutral stability curves (dashed blue) for specific modes represent where the equilibrium is neutrally stable to the given mode, but may already be unstable to other mode(s). The number next to each curve provides the corresponding Fourier mode(s). For each neutral stability curve, the set with \(F<0\) is the blue shaded region above the curve. Darker blue shaded regions indicate instability to multiple Fourier modes. Each bifurcation curve has two analytically computable asymptotes (solid red) which are described in Theorem \ref{thm: Asymptotes}. The solid black line represents the boundary of \(\mathcal{P}\) and the gray shaded region outside of \(\mathcal{P}\) are the parameter values for which the equilibrium is always stable. \textbf{(\subref{fig: h4N8 Bifurcation})} For \(h=4\) and \(N=8\), only the \(k=\pm 1\) modes are unstable. \textbf{(\subref{fig: h4N23 Bifurcation})} For \(h=4\) and \(N=23\), four modes are unstable. When \(b\) is small, the \(k=\pm 2\) modes lose stability first, but when \(b\) is larger it is the \(k=\pm 3\) modes. \textbf{(\subref{fig: h11N23 Bifurcation})} For \(h=11\), and \(N=23\), four modes are unstable. The \(k=\pm 1\) modes always lose stability before \(k=\pm 3\). The asymptotes are omitted from this figure because they are visually indistinguishable from the bifurcation curve itself.}
\label{fig: Bifurcation no asymptotes}
\end{figure}

In summary, for \(h\geq 3\) fixed and \(N\) chosen sufficiently large, \((b,d)\) may be chosen to produce a spatially periodic instability. This is not possible when \(h<3\) or \((b,d)\notin\mathcal{P}\). This means that the condition that melanophores die faster than they regrow in isolation is still required and that there is a minimum projection length for stripes to form. This condition that projections have length at least \(3\) times the cell diameter is the same bound on projection length that is observed in \cite{Hamada}. As with Theorem \ref{thm: Discrete Bifurcation}, the wavelength of these stripes has the same well-defined bounds which agree with nature. Additionally, the waves that first become unstable may be different for different regions of the \(bd\)-plane. Stripe widths in zebrafish tend to be close to double the length of projections \cite{Hamada}, and empirical evidence suggests that this occurs when the birth and death rate of melanophores are close (see Figure \ref{fig: Bifurcation no asymptotes}). So, the model is most similar to real fish when the in-isolation death rate of melanophores is greater than the birth rate by a small amount.

Figure \ref{fig: Bifurcation no asymptotes} shows the curves \(F=0\) in the \(bd\)-plane (as opposed to the \(xy\)-plane in Figures \ref{fig: h3N6 Cheb plots} and \ref{fig: b3d10 Cheb plots}). If \(h\) and \(N\) are sufficiently large and \(k\) is chosen appropriately, the level set \(F(x_{k},T_{h}(x_{k}),b,d)=0\) lies within the set \(\mathcal{P}\), bounded on the left by a vertical asymptote and bounded below and to the right by a slant asymptote (see Theorem \ref{thm: Asymptotes}). For each \(k\), the set \(F<0\) lies above this curve. In Figure \ref{fig: h4N8 Bifurcation}, there is exactly one pair of conjugate modes, \(k=\pm 1\), which become unstable, so the \(F(x_{1},T_{4}(x_{1}),b,d)=0\) level set is exactly the bifurcation curve on which the equilibrium loses stability. This is due to the small size of \(N\) relative to \(h\). If \(N\) is large enough, there will be multiple Fourier modes which become unstable as in Figures \ref{fig: h4N23 Bifurcation} and \ref{fig: h11N23 Bifurcation}. 

In the case of Figure \ref{fig: h4N23 Bifurcation}, the unstable modes are \(k=\pm 2,\pm 3\) and neither unstable set contains the other. For example, near \((b,d)=(1,6)\), only the \(k=\pm 2\) modes are unstable and near \((b,d)=(4,10)\) only the \(k=\pm 3\) modes are unstable. The sizes of these regions are small relative to the region where both modes are unstable. In this overlapping region, the dominant mode in the steady state is selected through a non-linear interaction, meaning that the linear analysis in this paper can only serve as a heuristic for estimating the modes which are more likely to dominate. Empirically, there is a trend that, as \(k\) increases within the range defined in inequalities \eqref{Bounds1} and \eqref{Bounds2}, the vertical asymptote moves further right and the slant asymptote becomes less steep. This indicates that if cells die much faster than they are created, then the resulting stripes will be wider and lesser in number. If cells are created and die at similar rates, the resulting stripes will be narrower and greater in number.

In the case of Figure \ref{fig: h11N23 Bifurcation}, the unstable modes are \(k=\pm 1, \pm 3\) and the \(k=\pm 1\) unstable set entirely contains the \(k=\pm 3\) set. This qualitative difference is directly related to the intuition developed in the discussion of Figure \ref{fig: b3d10 Cheb plots}. Figure \ref{fig: h11N23b3d10 Cheb plot} shows the \(F=0\) level set in the \(xy\)-plane for the same \(h\) and \(N\) values, with \((b,d)=(3,10)\). The point corresponding to the \(k=\pm 1\) modes lies between the rightmost minimum and rightmost root of \(y=T_{h}(x)\), whereas the \(k=\pm 3\) modes do not. This is the geometric intuition behind the bounds in \eqref{Bounds2} which indicate that the \(k=\pm 3\) modes cannot be the first to lose stability. The containment of the \(k=\pm 3\) unstable set within the \(k=\pm 1\) unstable set is the corresponding geometric phenomenon in the \(bd\)-plane. 

Figure \ref{fig: Bifurcation no asymptotes} also demonstrates that, as \(h\) increases, the set of parameter values where the equilibrium is unstable approaches the totality of the set \(\mathcal{P}\). For any fixed \(h\) and \(N\), as \(b\) increases the gap between the boundary of \(\mathcal{P}\) and the actual bifurcation curve for the equilibrium grows due to the slope of the slant asymptotes being strictly greater that \(1\). This means that if \(h\) is large and \(b\) is relatively small, the actual regime where the equilibrium loses stability is approximately the same as \(\mathcal{P}\), with the accuracy improving as \(h\) and \(N\) increase. This means that for large \(h\) the onset of stripe formation occurs at approximately the same rate parameter values. The primary differentiating factor in such cases will be the wavelength of the generated stripes. The bounds on the bifurcating Fourier modes imply that the width of observed stripes is roughly proportional to \(h\). This predicts that a zebrafish which produces longer melanophore stripes on average will display wider and potentially fewer stripes.

Theorems \ref{thm: Discrete Bifurcation} and \ref{thm: Continuous Bifurcation} both include conditions of \(N\) being sufficiently large for the conclusions to hold. For a model with applications for fish of a bounded size, it is important to also consider the question of what is the smallest value of \(N\) where instability may occur. This smallest value is \(N=6\). This can be shown by considering a special case in which taking \(N\) sufficiently large is not required for the proofs of the main theorems. As mentioned in the discussion of Figures \ref{fig: h3N6 Cheb plots} and \ref{fig: b3d10 Cheb plots}, in the proofs of these theorems the curve \(y=T_{h}(x)\) is shown to enter the region where \(F<0\) and then \(N\) is increased to guarantee that the points \((x,y)=(x_{k},T_{h}(x_{k}))\) are sufficiently dense in the curve. However, for the special case \(N=2h\), the point \(x_{\pm 1}\) is exactly at the rightmost local minima of the \(y=T_{h}(x)\) curve, as can be seen in Figure \ref{fig: h3N6 Cheb plots}. This eliminates the need for \(N\) to be sufficiently large because the curve \(y=T_{h}(x)\) is shown to intersect the region \(F<0\) by studying when this local minimum is in the region \(F<0\). So, if \(N=2h\) and \(h\geq 3\), the equilibrium \(\left(\tfrac{d}{b+d},\tfrac{b}{b+2d}\right)\) will be unstable for some \((b,d)\). This means that for model \eqref{1D ODE}, \(N=6\) is the absolute minimal case for instability to occur. This minimum size is sufficiently small that any actual fish being studied will be many times larger; the results are applicable at the physically realistic scale.

\begin{figure}[htbp]
\begin{subfigure}[t]{0.49\textwidth}
    \centering
    \includegraphics[width=0.9\linewidth]{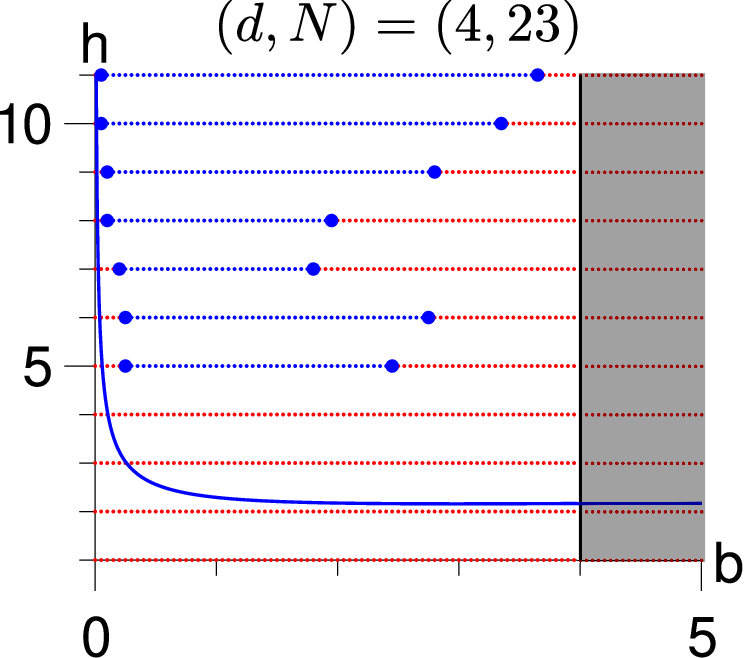}
    \subcaption{}
    \label{fig: d4N23 bhBifurcationPlot}
\end{subfigure}
\begin{subfigure}[t]{0.49\textwidth}
    \centering
    \includegraphics[width=0.9\linewidth]{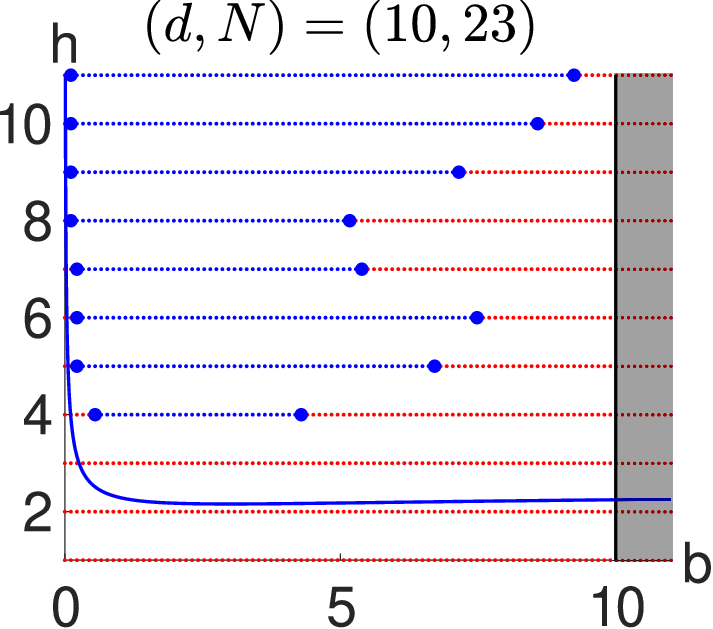}
    \subcaption{}
    \label{fig: d10N23 bhBifurcationPlot}
\end{subfigure}

\begin{subfigure}[t]{0.49\textwidth}
    \centering
    \includegraphics[width=0.9\linewidth]{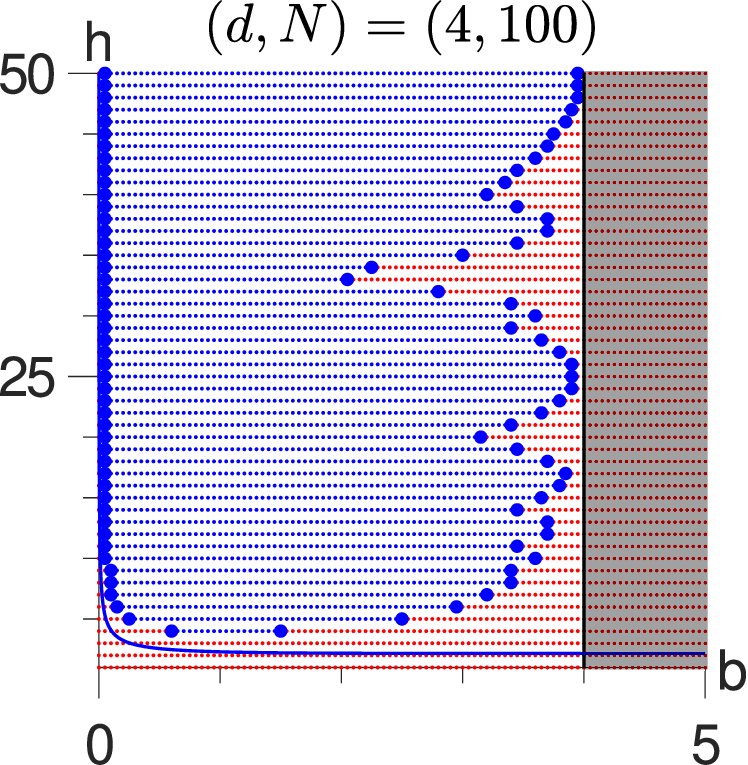}
    \subcaption{}
    \label{fig: d4N100 bhBifurcationPlot}
\end{subfigure}
\begin{subfigure}[t]{0.49\textwidth}
    \centering
    \includegraphics[width=0.9\linewidth]{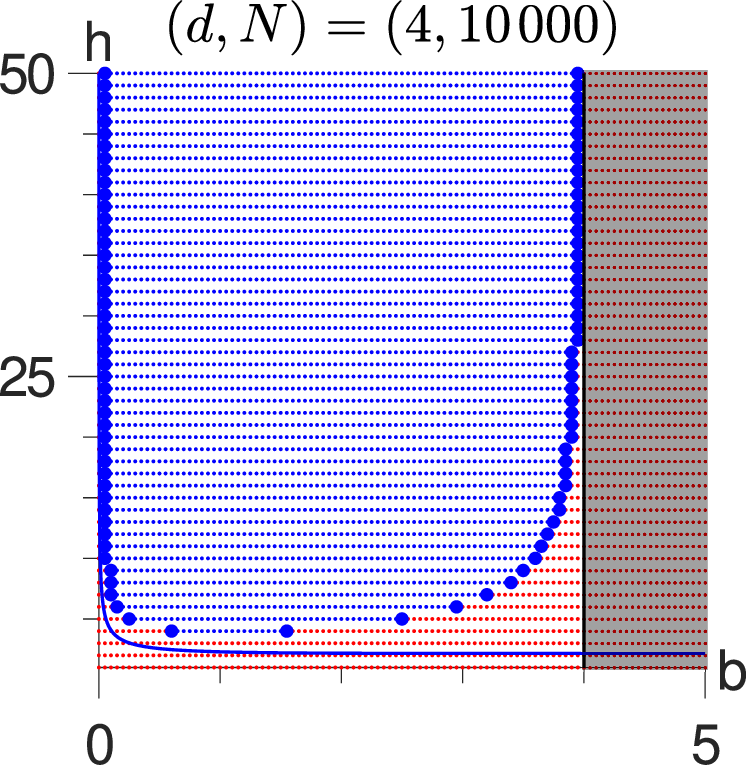}
    \subcaption{}
    \label{fig: d4LargeN bhBifurcationPlot}
\end{subfigure}
\caption{Scatter plots of linear instability behavior in the \(bh\)-plane for fixed \(d\) values. Each sample point is colored blue if the homogeneous state is unstable and red if it is stable The PDE model predicts instability for all \((b,d)\) above the PDE neutral stability curve (blue line). The largest and smallest unstable values are marked with larger dots. The shaded gray region with solid black boundary is the compliment of \(\mathcal{P}\) in each plot, meaning that the equilibrium is always linearly stable in this region. For panels (\subref{fig: d4N23 bhBifurcationPlot})-(\subref{fig: d4LargeN bhBifurcationPlot}), the plots range over all values of \(h\) for the given \(N\). \textbf{(\subref{fig: d4N23 bhBifurcationPlot})} For \(d=4\) and \(N=23\), the region of instability satisfies \(0< b < 5\) and \(5\leq h\). \textbf{(\subref{fig: d10N23 bhBifurcationPlot})} For \(d=10\) and \(N=23\), the qualitative shape is similar to the \(d=4\) case, but now there is an additional interval at the \(h=4\) level and all intervals extend to larger values of \(b\). \textbf{(\subref{fig: d4N100 bhBifurcationPlot})} For \(d=4\) and \(N=100\), the instability region again includes an interval at \(h=4\) and is bounded by \(b=5\), but the intervals generally extend closer to this boundary. The rich structure of the right boundary demonstrates the value of performing LSA for all values of \(h\) and \(N\). \textbf{(\subref{fig: d4LargeN bhBifurcationPlot})} For \(d=4\) and \(N=10000\), the left boundary of these intervals are qualitatively similar to the small \(N\) cases, but the right boundary maintains none of the fine structure observed in the \(N=100\) case.}
\label{fig: bhBifurcationPlots}
\end{figure}

The ability to compute linear stability for all combinations of \(h\) and \(N\) is highlighted in Figure \ref{fig: bhBifurcationPlots}. In Figure \ref{fig: bhBifurcationPlots}, a number of fixed \(d\) and \(N\) values are selected. For each of these selected values and all \(0\leq k \leq \hfrac{N}{2}\), \(100\) \(b\) values are sampled and \(F(x_{k},T_{h}(x_{k}),b,d)\) is evaluated. The \(b\) values where the equilibrium \(\left(\tfrac{d}{b+d},\tfrac{b}{b+2d}\right)\) is unstable are the union over all \(k\) of the \(b\) values where each \(k\) is unstable. Figure \ref{fig: bhBifurcationPlots} plots all sample values colored by the linear stability of the homogeneous equilibrium. This allows a comparison with the bifurcation curves in Figures \ref{fig: Bifurcation no asymptotes} and \ref{fig: h4N23vsLargeNBifurcation}. The finite \(N\) curves are computed using a pseudo-arclength continuation of the set \(\{F(x_{k},T_{h}(x_{k}),b,d)=0\}\), while the \(N\to\infty\) curve uses continuation of the set \(\{F(x,T_{h}(x),b,d)=0,\tfrac{d}{dx}F(x,T_{h}(x),b,d)=0\}\). The width of the region of instability in Figure \ref{fig: h11N23 Bifurcation} at the \(d=4\) and \(d=10\) levels agree well with the regions in Figures \ref{fig: d4N23 bhBifurcationPlot} and \ref{fig: d10N23 bhBifurcationPlot} at the \(h=11\) level.

By looking at the regions where the base state is unstable in the \(bh\)-plane, we see the complicated relationship between the parameters and the stability properties. For \(d\) and \(N\) fixed, predicting the minimum and maximum \(b\) values where Turing instability begins as a function of \(h\) is non-trivial. The maximum values in particular vary widely and are neither monotone nor periodic in \(h\). Increasing \(d\) has the effect of making the instability regions larger, but remaining qualitatively similar. Increasing \(N\) increases the complexity of the boundary and makes estimation even less promising. When \(N>>h\), the right boundary appears to be increasing in \(h\), which agrees with the expectations for the \(N\to\infty\) limit. For accurate descriptions of the linear stability of this type of coupled ODE system, it is necessary to perform the analysis for all parameter values, and not just the large \(N\) limits.

\subsection{Numerical Simulations}

\begin{figure}[htbp]
\begin{subfigure}[t]{0.98\textwidth}
    \centering
    \includegraphics[width=0.5\linewidth]{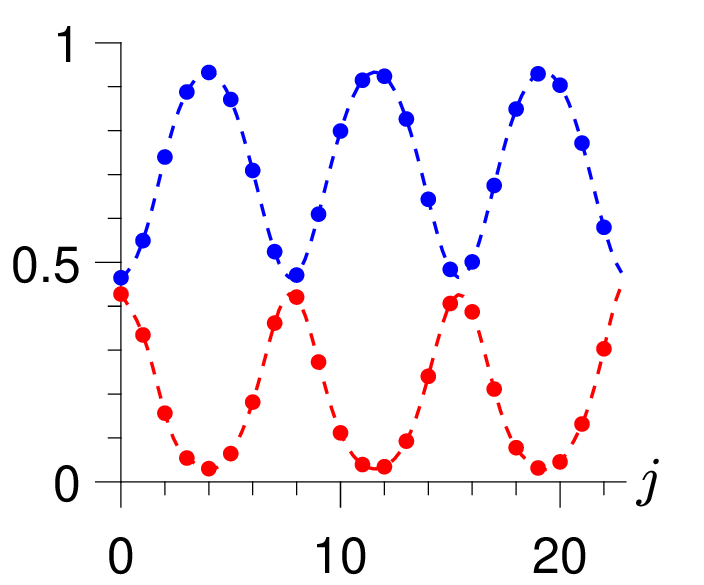}
    \subcaption{}
    \label{fig: h4N23b4d16 Final X}
\end{subfigure}

\begin{subfigure}[t]{0.49\textwidth}
    \centering
    \includegraphics[width=0.9\linewidth]{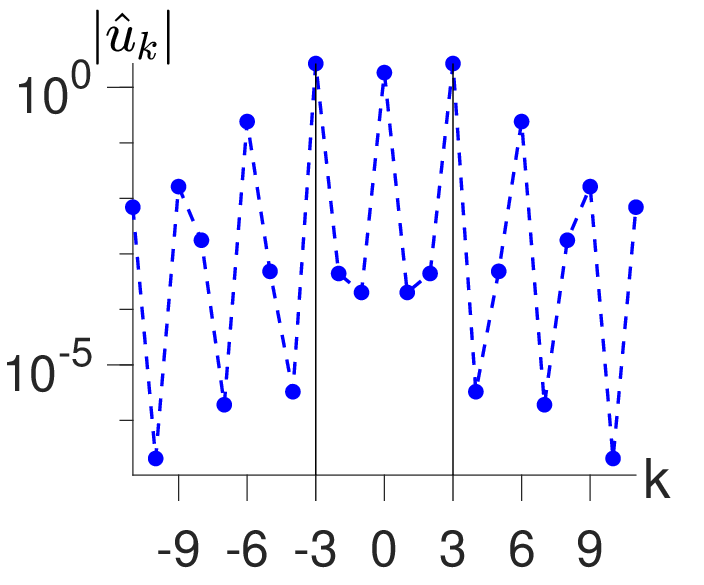}
    \subcaption{}
    \label{fig: h4N23b4d16 Final X FFT}
\end{subfigure}
\begin{subfigure}[t]{0.49\textwidth}
    \centering
    \includegraphics[width=0.9\linewidth]{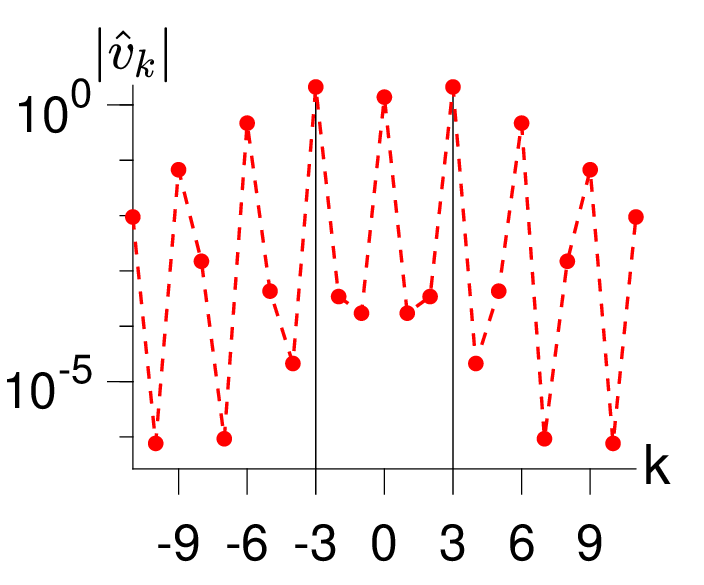}
    \subcaption{}
    \label{fig: h4N23b4d16 Final M FFT}
\end{subfigure}
    \caption{The numerical steady state for a simulation with parameters \(h=4\), \(N=23\), \(b=4\), and \(d=16\). \textbf{(\subref{fig: h4N23b4d16 Final X})} The steady state of the xanthophore variables (blue dots), melanophore variables (red dots) and the smooth interpolation of the Fourier spectrum of each (dashed lines). \textbf{(\subref{fig: h4N23b4d16 Final X FFT}-\subref{fig: h4N23b4d16 Final M FFT})} The magnitudes of the Fourier space representation of the steady state from \textbf{(\subref{fig: h4N23b4d16 Final X})} expressed as deviation from the homogeneous equilibrium on a logarithmic scale. Figure \textbf{(\subref{fig: h4N23b4d16 Final X FFT})} represents xanthophores and \textbf{(\subref{fig: h4N23b4d16 Final M FFT})} represents melanophores. The content of the \(0\) modes corresponds to a deviation of the average values by approximately \(-0.07\) for xanthophores and \(0.06\) for melanophores.}
    \label{fig: h4N23b4d16 Steady State}
\end{figure}

\begin{figure}[htbp]
\begin{subfigure}[t]{0.49\textwidth}
    \centering
    \includegraphics[width=0.99\textwidth]{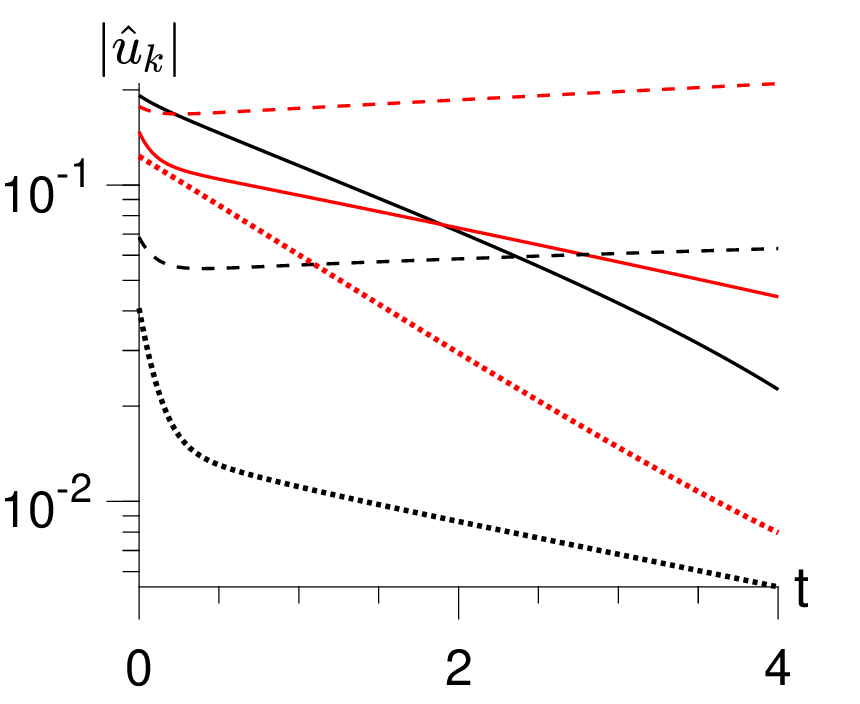}
    \subcaption{}
    \label{fig: h4N23 Super Short Time}
\end{subfigure}
\begin{subfigure}[t]{0.49\textwidth}
    \centering
    \includegraphics[width=0.99\textwidth]{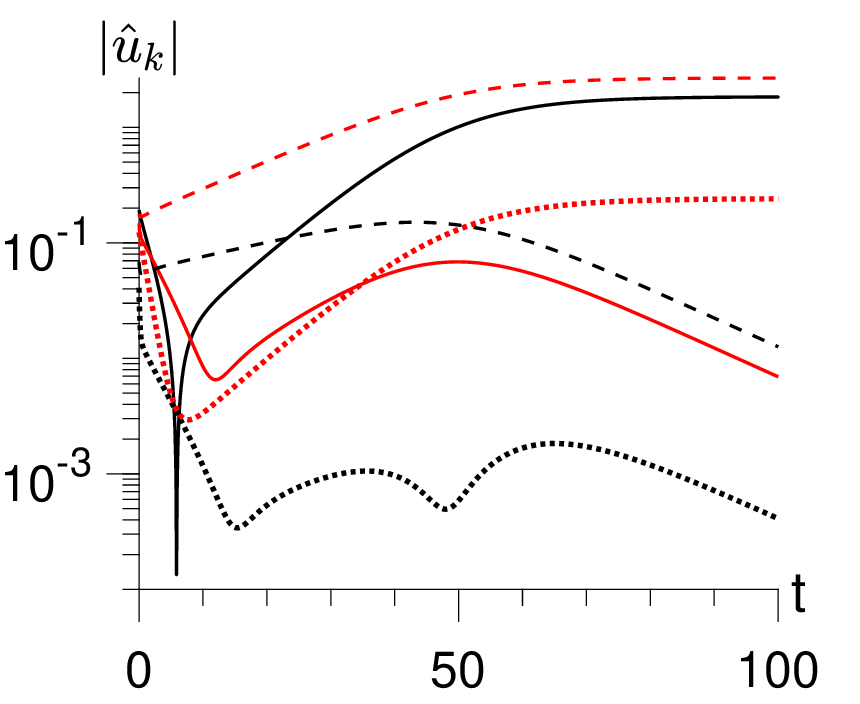}
    \subcaption{}
    \label{fig: h4N23 Short Time}
\end{subfigure}

\begin{subfigure}[t]{0.98\textwidth}
    \centering
    \includegraphics[width=0.495\textwidth]{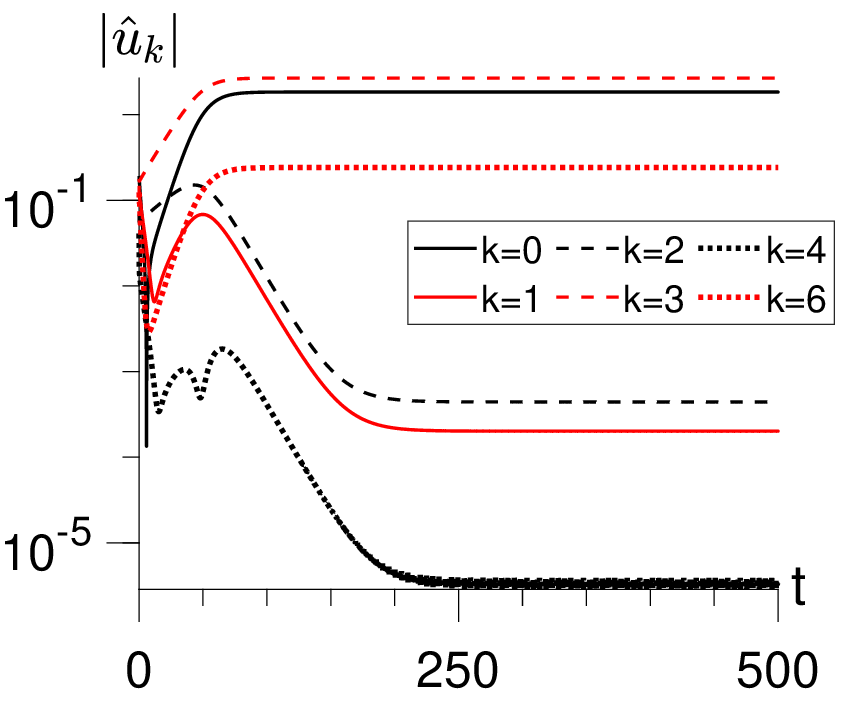}
    \subcaption{}
    \label{fig: h4N23 Long Time}
\end{subfigure}
    \caption{The time series data of the same simulation as Figure \ref{fig: h4N23b4d16 Steady State} for select Fourier modes on a logarithmic scale. \textbf{(\subref{fig: h4N23 Super Short Time})} Short-time activity for the selected modes is dominated by the linear dynamics. The \(k=2,3\) modes display saddle dynamics while all other modes display sink dynamics. \textbf{(\subref{fig: h4N23 Short Time})} Intermediate-time activity of selected modes displays the effects of non-linear coupling on the transient dynamics. Near \(t=50\), the \(k=3\) mode is selected as the dominant mode in the steady state. \textbf{(\subref{fig: h4N23 Long Time})} Long-time dynamics are determined by stable steady states. All values are near numerical equilibrium by \(t=250\) and do not change significantly by \(t=500\).}
    \label{fig: h4N23 Time Series}
\end{figure}

\begin{figure}[htbp]
\begin{subfigure}[t]{0.49\textwidth}
    \centering
    \includegraphics[width=0.98\textwidth]{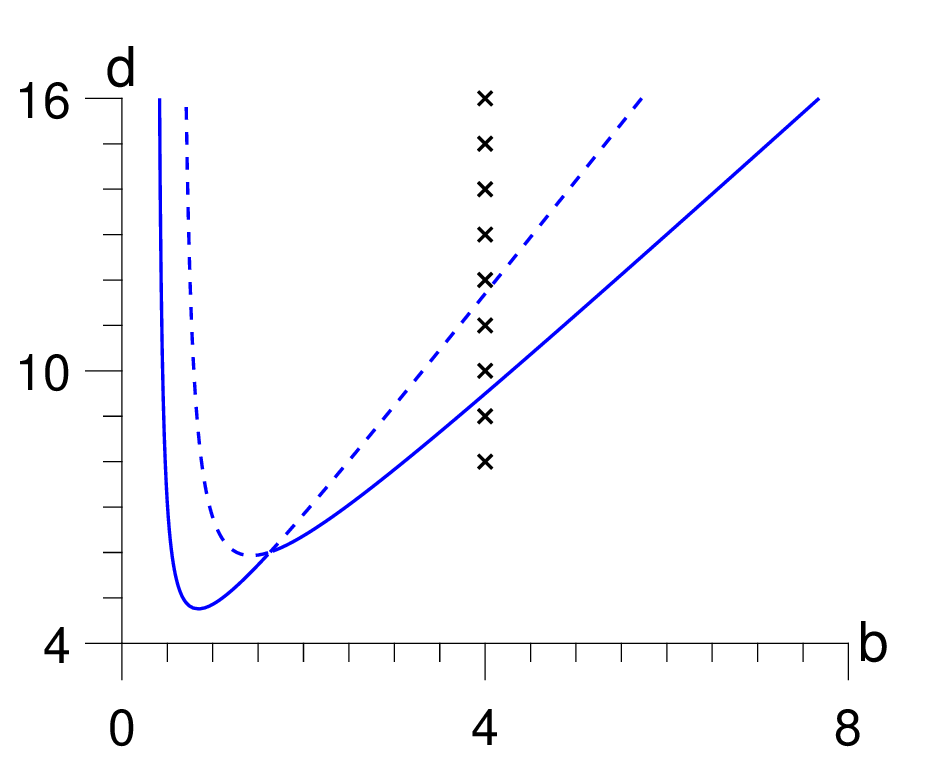}
    \subcaption{}
    \label{fig: h4N23 Bifurcation with Samples}
\end{subfigure}
\begin{subfigure}[t]{0.49\textwidth}
    \centering
    \includegraphics[width=0.98\textwidth]{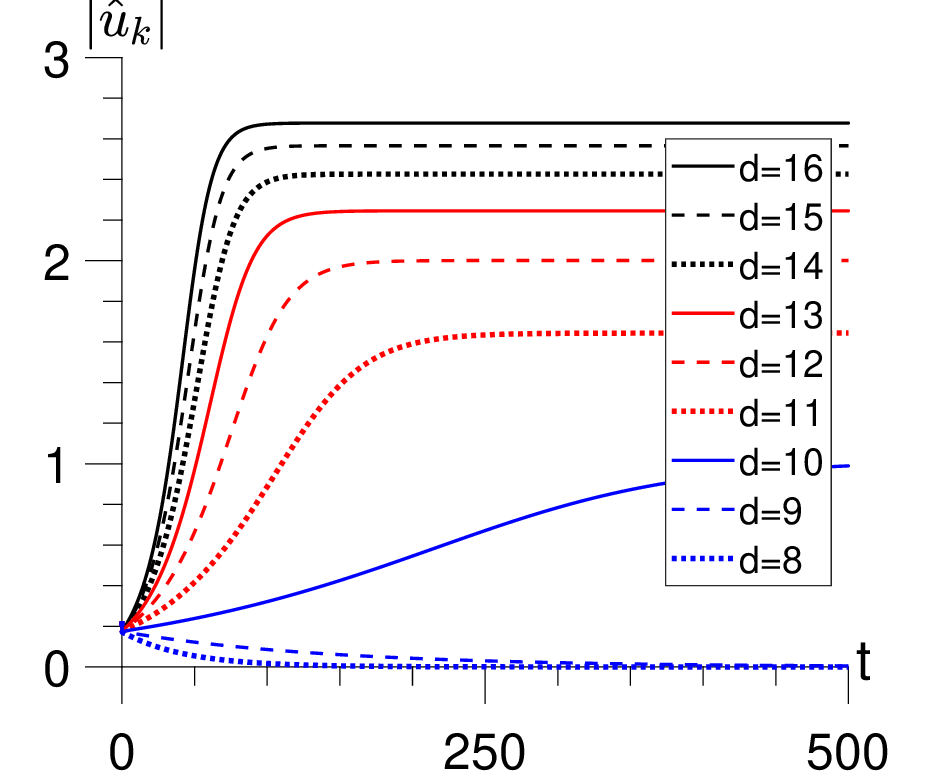}
    \subcaption{}
    \label{fig: h4N23b4 Time Series}
\end{subfigure}
\caption{Time series data for simulations with differing parameter values. For all simulations, \(h=4\), \(N=23\), and \(b=4\) (\((h\) and \(N\) are the same as in Figure \ref{fig: h4N23 Bifurcation}). \(d\) was sampled between \(8\) and \(16\) for comparison. In all simulations, the dominant mode at steady state was \(k=3\). \textbf{(\subref{fig: h4N23 Bifurcation with Samples})} The selected parameter values (black crosses) relative to the bifurcation curve (solid blue) and neutral stability curves (dashed blue) for specific modes. \textbf{(\subref{fig: h4N23b4 Time Series})} Time series data of the dominant modes in the steady state. Solution curves appear in the same order presented in the legend. Parameter values further into the unstable region experience faster exponential growth and larger equilibrium values. Parameter values on the stable side of the bifurcation curve experience exponential decay, as predicted by the linear analysis.}
\label{fig: h4N23b4 d Samples}
\end{figure}

Using MATLAB's ode89 program, we can simulate the system and compare the numerical results to the theoretical predictions. All simulations were initialized with random initial data distributed uniformly within a \(\pm 0.05\) window of the homogeneous equilibrium. Figures \ref{fig: h4N23b4d16 Steady State} and \ref{fig: h4N23 Time Series} depict the steady state and time series data respectively of a representative simulation. Figure \ref{fig: h4N23b4 d Samples} depicts time series data for a set of parameters values on either side of the bifurcation curve in Figure \ref{fig: h4N23 Bifurcation}.

Figure \ref{fig: h4N23b4d16 Steady State} depicts the numerical steady state for a simulation with \(h=4\), \(N=23\), \(b=4\), and \(d=16\). For these parameter values, the linear stability theory states that there should be instability in the \(k=\pm 2,\pm 3\) modes. In parameter space, these values are closer to the neutral stability curve of \(k=\pm 2\) than \(k=\pm 3\). Heuristically, one may expect the \(k=\pm 3\) modes to experience faster exponential growth than the \(k=\pm 2\) modes and be more likely to be the dominant mode in the steady state. While non-rigorous, this heuristic does correctly predict the dominant mode. In all simulations the melanophore expected values exhibited oscillations that were out of phase with the xanthophore, in agreement with theoretical results and observations in nature.

The Fourier space representations of the solution in Figures \ref{fig: h4N23b4d16 Final X FFT} and \ref{fig: h4N23b4d16 Final M FFT} express the difference between the steady state and the initial homogeneous state; the non-zero content of the \(k=0\) mode represents a difference in the average value between the two states. The Fourier space representation also highlights an important phenomenon in the study of Turing patterns on a coupled ODE system. Because \(N\) and the dominant value of \(k\) in the steady state are coprime, every mode is a resonant mode of the critical mode. In the presented example, the \(k=\pm 6, \pm 9\) are double and triple the dominant mode respectively. The \(k=\pm 11\) modes are \(4\) times the critical mode when considered mod \(23\); explicitly, \(\pm 11 \equiv 4(\mp3) \mod 23\). The zig-zag structure in the plot is actually an artifact of the higher resonance modes wrapping around at the boundary due to the periodic nature of the DFT. There is no similar phenomenon in the Fourier space representation of a periodic PDE because there is no way to express the positive modes as multiples of negative modes in the way that we can in the discrete case.

In Figure \ref{fig: h4N23 Time Series}, time series data demonstrates the three time regimes where different types of behavior occur. The subfigures depict a short time scale dominated by linear dynamics, an intermediate time dominated by non-linear effects, and a long time dominated by equilibrium stability respectively. The modes \(k=0,1,2,3,4,6\) were chosen to be representative of different behaviors among different modes. The dominant equilibrium mode is \(k=3\), the \(k=0,6\) modes are low order resonance modes of the dominant mode, \(k=2\) is a high order resonance mode which is also linearly unstable, and modes \(k=1,4\) are high order resonance modes of \(k=3\) which are also excited by interactions of the \(k=2\) and \(k=3\) modes.

In the short time, all modes exhibit a short period of fast exponential decay corresponding to the most negative eigenvalue. Then, modes \(k=2,3\) experience slow exponential growth from the positive eigenvalues predicted in the LSA. The other modes experience a period of slow exponential decay from the less negative eigenvalues. In the intermediate time, the non-linear coupling causes excitation in the modes which are sums of the unstable modes. The \(k=1\) mode is excited by the interaction of the \(k=-2\) and \(k=3\) modes, for example. In this simulation, the random perturbations increased the initial average value of xanthophores relative to the homogeneous state, but the periodic equilibrium state has an average value approximately \(0.07\) lower than the equilibrium state. This leads to the downward spike in the \(k=0\) curve near \(t=6\) as the average value passes through the homogeneous state. In the long time scale, all modes approach constant values. Across simulations, the numerical error near amplitude \(10^{-6}\) becomes more significant, even for other ode solvers.

In Figure \ref{fig: h4N23b4 d Samples}, we examine the effect of changing \(d\) on the evolution of the system while \(b=4\), \(h=4\) and \(N=23\) are held constant. Figure \ref{fig: h4N23 Bifurcation with Samples} depicts the same bifurcation curve as Figure \ref{fig: h4N23 Bifurcation} with the specific values used in the simulations marked for quick comparison with the expected linear behavior. Figure \ref{fig: h4N23b4 Time Series} shows the time evolution of the dominant mode at equilibrium (\(k=\pm 3\)) for each sample. For \(d=8,9\), the equilibrium is linearly stable and the dominant modes decay to \(0\) with \(d=8\) decaying faster than \(d=9\). For the cases \(d=10,...,16\), we see the expected exponential growth of the dominant modes before their eventual saturation at steady state. As \(d\) increases and the parameter value moves further into the region of instability, the rate of exponential growth and the amplitude at saturation both increase. For larger \(d\), the increase in saturation amplitude is less significant. This is analogous to the behavior of supercritical Turing Bifurcations in PDEs.

\subsection{Other Results}
Theorem \ref{thm: Asymptotes} and Corollary \ref{cor: Asymptotes} provide results regarding the behavior of the neutral stability curves of individual Fourier modes and bifurcation curves respectively. Namely, these curves approach asymptotes in both directions and the formulas for these asymptotes can be calculated explicitly.

\begin{theorem}\label{thm: Asymptotes}

    For the value(s) of \(k\) in Theorem \ref{thm: Continuous Bifurcation} which are unstable, \(\mathcal{U}_{k}(h,N)\) is bounded on the left by a vertical asymptote and bounded below and on the right by a slant asymptote. These asymptotes are characterized by the following equations:
    \begin{subequations}
    \begin{align}
        b=&\frac{2(x_{k}-1)}{f_{2}(x_{k})}\label{Finite N Vertical Asymptote};\\
        d=&\tan(\theta)b+\frac{(x_{k}-1)\left((x_{k}+3)\sin(2\theta)-\cos(2\theta)+3\right)}{2\cos^{2}(\theta)\left(f_{1}(x_{k})\cos(2\theta)+(f_{2}(x_{k})-1)\sin(2\theta)\right)}\label{Finite N slant Asymptote},\\
        2\cot(\theta)=&f_{1}(x_{k})+\sqrt{T_{h}^{2}(x_{k})-6x_{k}T_{h}(x_{k})+2T_{h}(x_{k})+(x_{k}+1)^{2}}.
    \end{align}
    where \(f_{i}(x)\) are defined as
    \begin{align*}
        f_{1}(x)=&-x+T_{h}(x)+3,\\
        f_{2}(x)=&(x+1)T_{h}(x)-2(x-1).
    \end{align*}
    \end{subequations}
    For all \(k\) which are unstable, \(\hfrac{\pi}{4}<\theta<\hfrac{\pi}{2}\).
\end{theorem}

\begin{corollary}\label{cor: Asymptotes}

    For \(h\) and \(N\) satisfying the hypothesis of Theorem \ref{thm: Continuous Bifurcation}, the bifurcation curve in the \(bd\)-plane, characterized by
    \begin{align}
        \partial\left(\bigcup_{k=0}^{N-1}\mathcal{U}_{k}(h,N)\right),
    \end{align}
    is bounded on the left by a vertical asymptote and eventually bounded below and on the right by a slant asymptote. These asymptotes are characterized by the following equations:
    \begin{subequations}
        \begin{align}
            b=&\inf_{k}\left(\frac{-2(1-x_{k})}{(x_{k}+1)T_{h}(x_{k})+2(1-x_{k})}\right);\\
            d=&\tan(\theta)b+\frac{(x_{k}-1)\left((x_{k}+3)\sin(2\theta)-\cos(2\theta)+3\right)}{2\cos^{2}(\theta)\left(f_{1}(x_{k})\cos(2\theta)+(f_{2}(x_{k})-1)\sin(2\theta)\right)},\\
            2\cot(\theta)=&\sup_{k}\left(f_{1}(x_{k})+\sqrt{T_{h}^{2}(x_{k})-6x_{k}T_{h}(x_{k})+2T_{h}(x_{k})+(x_{k}+1)^{2}}\right).
        \end{align}
        where the infimum and supremum are taken over the values of \(k\) which are unstable from Theorem \ref{thm: Continuous Bifurcation}.
    \end{subequations}
\end{corollary}
In short, the neutral stability curve for each Fourier mode has a vertical and slant asymptote for fixed \(h\) and \(N\). The vertical asymptote is always positive, but the asymptote becomes arbitrarily large for some \(k\) as \(N\) increases. The slope of the slant asymptote is always larger than \(1\) and becomes arbitrarily large under similar conditions. The asymptotes for the overall bifurcation curve are found by considering the extremal cases of the asymptotes for the individual Fourier modes. For the vertical asymptote, this is the asymptote with the minimal intercept. For the slant asymptote, this is the asymptote with the minimal slope. These asymptotes can be used to supplement the direct numerical computation of neutral stability curves. The curve can be computed near the origin and continued until it is sufficiently close to the asymptotes. These results also allow us to make the bounds on birth and death rate more precise for any fixed \(h\) and \(N\). For example, when \(h=4\) and \(N=23\) and \((b,d)\) is far from the origin, the homogeneous state is unstable when \(b>0.3741...\) and \(d>(1.8185...)b+1.8721...\). Without these asymptotes, the best estimate for the ratio of \(d\) and \(b\) that produced stripes with wavelengths double the length of a projection was "\(d\) is slightly larger than \(b\)"; we now know \(d\) should be approximately \(1.8186\) times \(b\).

We now present results analogous to Theorems \ref{thm: Continuous Bifurcation} and \ref{thm: Asymptotes} and Corollary \ref{cor: Asymptotes} which consider limiting behavior as \(N\to\infty\). One motivation for these results is to predict the behavior of the model in a large fish. While the model was designed for zebrafish, it is applicable to any creature which uses a similar mechanism for pattern formation, including some which may be much larger than zebrafish. A second motivation is to provide a point of comparison against the predictions of the continuous mean-field survival model PDE. In particular, these results highlight that the differences between the PDE model and ODE model are not explained by the difference between finite \(N\) and infinite limits.

\begin{corollary}\label{cor: Large N Bifurcation}

    Let \(x_{\phi}=\cos\left(2\pi \phi\right)\). \(\forall h\geq 3\), \(\exists\) an interval \(I\subset \left(\hfrac{1}{4h},\hfrac{1}{2h}\right)\) depending only on \(h\) such that
    \begin{align}
        \lim_{N\to \infty}\left(\bigcup_{k=0}^{N-1}\mathcal{U}_{k}(h,N)\right)=\bigcup_{\phi\in I}\mathcal{U}_{\phi}(h),
    \end{align}
    where \(\mathcal{U}_{\phi}(h)=\left\{(b,d)\in\R_{+}^{2}:F(x_{\phi},T_{h}(x_{\phi}),b,d)<0\right\}\).
    
    The limiting bifurcation curve, defined by
    \begin{align}
        \partial\left(\bigcup_{\phi\in I}\mathcal{U}_{\phi}(h)\right),
    \end{align}
    is equivalent to the curve
    \begin{align}
        \left\{(b,d)=\R_{+}^{2}:F(x,T_{h}(x),b,d)=0,\partial_{x}F(x,T_{h}(x),b,d)=0\right\}.
    \end{align}
    This curve is bounded on the left by a vertical asymptote and eventually bounded below and on the right by a slant asymptote. These asymptotes are characterized by the following equations:
    \begin{subequations}
        \begin{align}
            b=&\inf_{\phi\in I}\left(\frac{-2(1-x_{\phi})}{(x_{\phi}+1)T_{h}(x_{\phi})+2(1-x_{\phi})}\right);\\
            b=&\tan(\theta_{*})d+C,\\
            C=&\frac{(x_{\phi_{*}}-1)\left((x_{\phi_{*}}+3)\sin(2\theta_{*})-\cos(2\theta_{*})+3\right)}{2\cos^{2}(\theta_{*})\left(f_{1}(x_{\phi_{*}})\cos(2\theta_{*})+(f_{2}(x_{\phi_{*}})-1)\sin(2\theta_{*})\right)},\\
            2\cot(\theta_{*})=&\sup_{\phi\in I}\left(f_{1}(x_{\phi})+\sqrt{T_{h}(x_{\phi})^{2}-6x_{\phi}T_{h}(x_{\phi})+2T_{h}(x_{\phi})+(x_{\phi}+1)^{2}}\right),\label{Large N Sup}
        \end{align}
    \end{subequations}
    where \(\phi_{*}\) is the value of \(\phi\) which produces the supremum in \eqref{Large N Sup}.
\end{corollary}

These \(N\to \infty\) results can be understood by considering limits with \(\hfrac{k}{N}\to\phi\in I\) as \(N\to \infty\). This corresponds to considering stripes with a fixed width as the total ring size increases. Under such conditions, \(x_{k}\to x_{\phi}\) and the analogues between the main theorems and Corollary \ref{cor: Large N Bifurcation} become more clear. The bounds on \(k\) in \eqref{Bounds2} imply bounds on the \(\phi\) values \(\hfrac{k}{N}\) could converge to and these bounds are exactly the bounds on the interval \(I\). The union, infimum, and supremum indexed over all unstable \(k\) in the finite \(N\) cases are now indexed by \(I\) in the \(N\to\infty\) case. These results are analogous to those in the large ring analysis performed by Turing in \cite{Turing}.

From a biological perspective, these results predict that even in large fish, stripe width should remain near a fixed chemical wavelength. Under the assumption that the pattern reaches equilibrium much faster than the fish grows, these results provide an estimate of how the pattern might evolve. Initially, the addition of only a few new cells to the ring will lead to a slight widening of stripes. Eventually, the width of stripes will be so much wider than the chemical wavelength that the pattern becomes unstable to the addition of a new stripe or stripes. Given the bounds on which modes are unstable, one might guess that this likely occurs before the stripes reach \(4\) times the length of melanophore projections. As the system tends towards equilibrium at this larger size, the existing stripes shrink and the new stripe grows until they are of similar size. The bounds on \(k\) suggest the new stripes will likely be at least double the length of projections.

\begin{figure}[htbp]
\begin{subfigure}[t]{0.98\textwidth}
    \centering
    \includegraphics[width=0.98\textwidth]{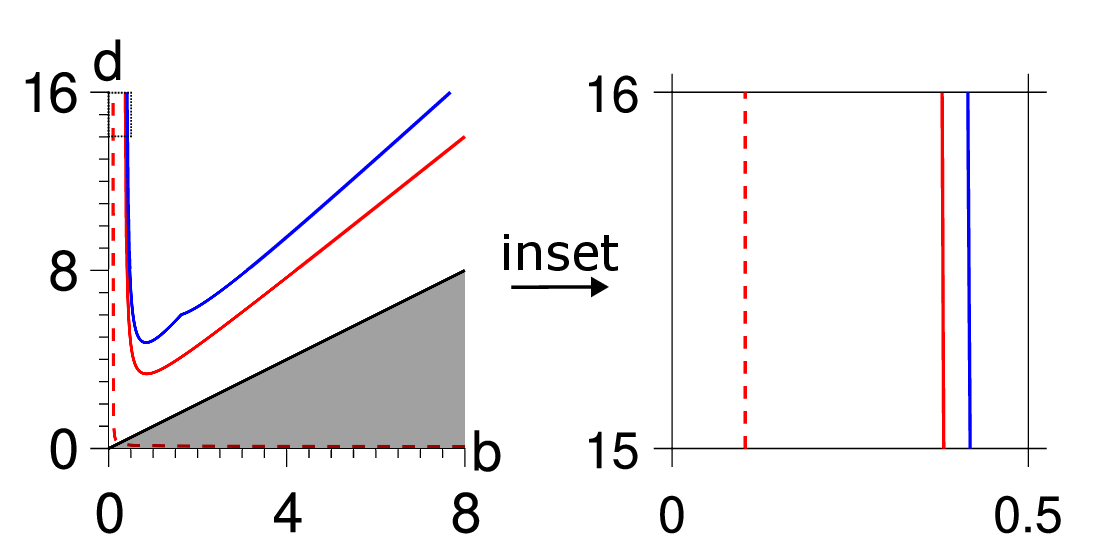}
    \label{fig: h4N23vsLargeNBifCurve}
\end{subfigure}
    \caption{The bifurcation curve (solid blue) for the ODE system with \(h=4\) and \(N=23\) compared to two different large population limits. The curve defined by \(F(x,T_{4}(x),b,d)=0\) and \(\partial_{x}F(x,T_{4}(x),b,d)=0\) (solid red) represents the \(N\to\infty\) limit predicted by the ODE analysis. The PDE bifurcation curve (dashed red) represents where the continuous mean-field survival model loses stability. For all three curves, the region of instability is the region above the curves. The gray shaded region with black boundary represents the region outside of \(\mathcal{P}\) for which the ODE system is always linearly stable. The limiting ODE curve has a similar shape to the \(N=23\) curve, including vertical and slant asymptotes and remaining in \(\mathcal{P}\). The PDE curve, however, is dissimilar in shape and even exits \(\mathcal{P}\). The PDE curve has a vertical asymptote, but the limiting ODE asymptote is closer to  that of the \(N=23\) ODE case as seen in the inset on the right.}
    \label{fig: h4N23vsLargeNBifurcation}
\end{figure}

From a mathematical perspective, the large \(N\) limit of the ODE system is similar to the finite \(N\) cases in ways that the PDE predictions are not. This is best demonstrated in Figures \ref{fig: bhBifurcationPlots} and \ref{fig: h4N23vsLargeNBifurcation}. As discussed previously, Figure \ref{fig: bhBifurcationPlots} shows bifurcation diagrams in the \(bh\)-plane for select values of \(d\) and \(N\). Particularly relevant are Figures \ref{fig: d4N100 bhBifurcationPlot} and \ref{fig: d4LargeN bhBifurcationPlot} depicting \(d=4\) and \(N=100\) and \(N=10000\) respectively. The bifurcation curve generated by the PDE has a reasonable vertical asymptote for small \(b\), but predicts instability for arbitrarily large values of \(b\) rather than only values in \(\mathcal{P}\). The \(N\to \infty\) ODE limit is more accurate in the sense that it indicates that \(b\) should be bounded in \(\mathcal{P}\). The large \(N\) ODE limit also predicts that the region of instability should eventually be intervals with widths increasing in \(h\), which agrees well with the \(N=10000\) case. However, the finite \(N\) case can still have fine structure to the upper bounds which is lost in the \(N\to \infty\) limit.

Figure \ref{fig: h4N23vsLargeNBifurcation} depicts bifurcation diagrams in the \(bd\)-plane for the PDE predictions, the \(N\to \infty\) ODE limit, and a finite \(N\) reference case. The \(N\to \infty\) limit corresponds to the curve where both \(F(x,T_{h}(x),b,d)=0\) and \(\tfrac{d}{dx}F(x,T_{h}(x),b,d)=0\). As with the previous figure, the difference between the PDE and large \(N\) ODE limit is significant. The PDE curve has vertical and horizontal asymptotes near the axes, predicting that almost all admissible parameter values should be unstable for \(h=4\). It again includes values outside of the region \(\mathcal{P}\), a major discrepancy from the ODE behavior. The large \(N\) ODE curve remains within the region \(\mathcal{P}\) and has a similar shape to the \(N=23\) case. As expected, the unstable region of the \(N\to \infty\) limit entirely contains the \(N=23\) region. The vertical asymptote of the ODE limit is much closer to the \(N=23\) asymptote than the PDE asymptote is. The slant asymptotes of the \(N=23\) and \(N\to \infty\) cases are significantly further from each other than the corresponding vertical asymptotes, indicating that ring size has less effect on the small \(b\) boundary than the large \(b\) boundary. The slope of the large \(N\) asymptote is generically less than any fixed \(N\), so the difference will become more pronounced for large values of \(b\). Similar to the improved bounds that the \(h=4\) and \(N=23\) asymptotes provided on the onset of instability, the asymptotes for \(h=4\) and \(N\to \infty\) provide improved bounds which are uniform for \(h=4\) and all values of \(N\). Specifically, \(b>0.3504...\) and \(d>(1.5961...)b+1.1900...\). So when \(h=4\), even for large \(N\), \(d\) must be at approximately \(1.5962\) times \(b\) to produce stripes with widths approximately double the length of projections. Overall, Figures \ref{fig: bhBifurcationPlots} and \ref{fig: h4N23vsLargeNBifurcation} demonstrate that the \(N\to \infty\) limit of the ODE system provides a significant improvement in the prediction of stability behavior relative to the PDE model, but also that the consideration of specific \(N\) values is still necessary for some details.

\section{Proofs of Main Results}

To prove the main results, we make use of the following lemma regarding useful properties of the function \(F\):
\begin{lemma}\label{lem: F Properties}

    \(F\) has the following properties:
    \begin{itemize}
        \item \(\forall (b,d)\in\mathcal{P}\), the set \(\mathcal{R}(b,d)=\left\{(x,y)\in[-1,1]^{2}:F(x,y,b,d)<0 \right\}\) is a neighborhood of \((x,y)=(1,-1)\) contained in the region defined by
        \begin{align}
            \left(1-\frac{b(b+2d)(d-b)}{b^{2}d+3bd^{2}+b^{2}+3bd+2d^{2}},1\right]\times\left[-1,\hfrac{-b}{d}\right)&\subseteq (0,1]\times[-1,0).
        \end{align}
        \item \(\forall (b,d)\in\mathcal{P}\), the set \(\left\{x\in\R:F(x,-1,b,d)<0\right\}\) contains the interval \((L,1]\) where
        \begin{align}
            L=&\frac{b^{3}+2b^{2}d+bd^{2}+b^{2}+4bd+2d^{2}}{b^{2}d+3bd^{2}+b^{2}+4bd+2d^{2}}.
        \end{align}
        The infimum of \(L\) over the set \(\mathcal{P}\) is \(\hfrac{1}{3}\).
        \item \(\forall C\leq 0\) the level sets \(F(x,y,b,d)=C\) in \([-1,1]^{2}\) can be expressed  with \(y\) a concave down, increasing function of \(x\).
        \item \(\forall (b,d)\notin\mathcal{P}\), \(\mathcal{R}(b,d)=\emptyset\).
    \end{itemize}
\end{lemma}
The region bounding \(\mathcal{R}(b,d)\) has the alternative characterization
\begin{align*}
    \left(\frac{b(b+d)^{2}+(b+d)(b+2d)}{b^{2}d+3bd^{2}+b^{2}+3bd+2d^{2}},1\right]\times\left[-1,\hfrac{-b}{d}\right)\subseteq (0,1]\times[-1,0).
\end{align*}
These two equivalent characterizations emphasize that the lower bound in the \(x\)-direction is strictly between \(0\) and \(1\) for all \((b,d)\in\mathcal{P}\). The bounding region is therefore both non-empty and contained in the set \((0,1]\times[-1,0)\). Proof of this lemma is technical and reserved for the appendix.
We can now prove Theorems \ref{thm: Discrete Bifurcation} and \ref{thm: Continuous Bifurcation}. 
\begin{proof}[Proof of Theorem \ref{thm: Discrete Bifurcation}]

    Recall that the Jacobian matrix of the linearization around this equilibrium has
    \begin{align*}
    \Tr\left(\hat{L}_{k}\right)=&\frac{-2(b+d)^{2}-(b+1)(b+2d)^{2}}{(b+d)(b+2d)},\\
    \Det\left(\hat{L}_{k}\right)=&\frac{F\left(x_{k},T_{h}(x_{k}),b,d\right)}{(b+d)(b+2d)},
    \end{align*}
    for all modes \(k\in\left\{0,...,N-1\right\}\). For admissible values of \((b,d)\), \(\Tr\left(\hat{L}_{k}\right)<0\) and \(\Det\left(\hat{L}_{k}\right)\) has the same sign as \(F\left(x_{k},T_{h}(x_{k}),b,d\right)\). So, there is linear instability if and only if \(\exists k\) with \(F\left(x_{k},T_{h}(x_{k}),b,d\right)<0\). By Lemma \ref{lem: F Properties}, the set \(\mathcal{R}(b,d)\) for which \(F(x,y,b,d)<0\) is non-empty exactly when \((b,d)\in\mathcal{P}\). Therefore, we may conclude that \((b,d)\notin\mathcal{P}\) implies that the equilibrium is completely linearly stable. This establishes the last statement of the theorem.
    
    To show the other results in the theorem, we first characterize when \(F(x,T_{h}(x),b,d)<0\) for \((b,d)\in\mathcal{P}\). For this, we utilize properties of Chebyshev polynomials of the first kind.
    \begin{itemize}
        \item The roots of \(T_{h}(x)\) occur at \(x=\cos\left(\hfrac{\pi(j+\hfrac{1}{2})}{h}\right)\) for \(j\in\left\{0,...,h-1\right\}\) and these roots are simple.
        \item The local extrema of \(T_{h}(x)\) occur at \(x=\cos\left(\hfrac{\pi j}{h}\right)\) for \(j\in\left\{1,...,h-1\right\}\).
        \item All local maxima of \(T_{h}(x)\) have value \(1\). All local minima of \(T_{h}(x)\) have value \(-1\).
        \item \(T_{h}(1)=1\).
    \end{itemize}
    The rightmost local minimum is \(x=\cos\left(\hfrac{\pi}{h}\right)\). By taking \(h\) sufficiently large, this local minimum can be made arbitrarily close to \(x=1\). For any fixed \((b,d)\in\mathcal{P}\), the set \(\mathcal{R}(b,d)\) contains some open rectangle \((1-\epsilon,1]\times[-1,-1+\delta)\). For \(h\) sufficiently large, the local minimum \(\left(\cos\left(\hfrac{\pi}{h}\right),-1\right)\) will be inside of this rectangle, guaranteeing that \(F(x,T_{h}(x),b,d)<0\) on some open interval of \(x\) values containing \(\cos\left(\hfrac{\pi}{h}\right)\). The fact that \(\mathcal{R}(b,d)\subseteq (0,1]\times[-1,0)\) also indicates this open interval is a subset of \(\left(\cos\left(\hfrac{3\pi}{2h}\right),\cos\left(\hfrac{\pi}{2h}\right)\right)\). We have now established that there exists an \(H_{1}\in \N\) such that \(h\geq H_{1}\) are all sufficiently large for \(F(x,T_{h}(x),b,d)\) to be negative on some open interval of \(x\) values. The existence of this interval is necessary but not sufficient for linear instability.
    
    The final step to demonstrating linear instability is to characterize when the intervals where \(F(x,T_{h}(x),b,d)<0\) contain points \(x_{k}\) for some \(k\in\left\{0,...,N-1\right\}\). Recalling that \(x_{k}=\cos\left(2\pi \hfrac{k}{N}\right)\), we derive the following bound on \(\lvert x_{k}-x_{k+1}\rvert\) using the sum-to-product trigonometric identities:
    \begin{align*}
        \lvert x_{k}-x_{k+1}\rvert=&\left\lvert \cos\left(2\pi \hfrac{k}{N}\right)-\cos\left(2\pi \hfrac{k+1}{N}\right)\right\rvert\leq 2\sin\left(\hfrac{\pi}{N}\right).
    \end{align*}
    Thus, by also making \(N\) sufficiently large, we are able to make \(\lvert x_{k}-x_{k+1}\rvert\) arbitrarily small. For any \((b,d)\in\mathcal{P}\) fixed and \(h\) sufficiently large, we can then choose \(N\) sufficiently large that \(\lvert x_{k}-x_{k+1}\rvert\) is smaller than the length of the interval of \(x\) values where \(F(x,T_{h}(x),b,d)<0\). We then know that \(\exists k\) such that \(F(x_{k},T_{h}(x_{k}),b,d)<0\) and conclude linear instability to appropriate perturbations. We can rule out the case \(k=0\) using the fact that \(T_{h}(1)=1\) and \(F(1,1,b,d)>0\). So, even if \(N\) is not large enough to establish the bounds on \(k\) proven below, it will always be the case that \(k\neq 0\) and so the instability will not be spatially homogeneous. This establishes the fact that the equilibrium is unstable to some spatially periodic perturbations under appropriate conditions.

    The characterization of the most linearly unstable waves makes use of the result from Lemma \ref{lem: F Properties} which states the level sets \(F=C\) for \(C\leq 0\) are concave down and increasing in \(x\). Because \(y=T_{h}(x)\) have all local minima equal to the global minima on \([-1,1]\), any line connecting the rightmost local minima to any point on \(y=T_{h}(x)\) left of the local minima must be non-increasing. So, the rightmost local minima must be on a more negative level set than any point on the curve to the left. We also know \(F(x,y,b,d)>0\) for \(y>0\) so all points on the curve \(y=T_{h}(x)\) right of the rightmost root must also be positive. So, the global minimum of \(F(x,T_{h}(x),b,d)\) must be between the rightmost local minimum and rightmost root of \(T_{h}(x)\). By taking \(N\) sufficiently large, \(x_{k}\) can be guaranteed to exist in the interval of values where \(F(x_{k},T_{h}(x_{k}),b,d)<F\left(\cos\left(\hfrac{\pi}{h}\right)),-1,b,d\right)\). Clearly, the value of \(F\) at any such \(x_{k}\) values is less than those of any \(x_{k}\) not in this interval. This interval is guaranteed to be inside of the interval between the rightmost local minimum and rightmost root, \(\left(\cos\left(\hfrac{\pi}{h}\right),\cos\left(\hfrac{\pi}{2h}\right)\right)\). The bounds in terms of \(k\) follow from the definition of \(x_{k}\) and appropriate rounding. This establishes the bounds on the values of \(k\) corresponding to the largest positive eigenvalue and concludes the proof.
\end{proof}

The proof of Theorem \ref{thm: Continuous Bifurcation} makes use of many of the same properties of \(F(x,T_{h}(x),b,d)\) and density arguments demonstrated in the proof of Theorem \ref{thm: Discrete Bifurcation}.
\begin{proof}[Proof of Theorem \ref{thm: Continuous Bifurcation}]

    For \(h\geq 3\), the rightmost local minimum of \(T_{h}(x)\), located at \(x=\cos\left(\hfrac{\pi}{h}\right)\), is greater than \(\hfrac{1}{3}\). By the established properties of \(\mathcal{R}(b,d)\) in Lemma \ref{lem: F Properties}, \(\left\{x\in\R:F(x,-1,b,d)<0\right\}\) contains an interval \((L,1]\) and \(\inf_{(b,d)\in\mathcal{P}}L=\hfrac{1}{3}\). Therefore, for \(h\geq 3\), \(F(\cos\left(\hfrac{\pi}{h}\right),-1,b,d)<0\) for some set of \((b,d)\) values. We now use this to prove the existence of some \(N_{3}\) with the desired properties via a proof by contradiction. We desire that \(N\geq N_{3}\) guarantee the existence of some admissible \((b,d)\) such that \(F(x_{k},T_{h}(x_{k}),b,d)<0\) for some \(k\neq 0\). For the sake of contradiction, we assume no such \(N_{3}\) exists; \(\forall M\in\N\), \(\exists N\geq M\) such that \(\forall (b,d)\) which are admissible and \(\forall k\), \(F(x_{k},T_{h}(x_{k}),b,d)\geq 0\). Therefore, for any admissible \((b,d)\), we can construct an increasing sequence \((N_{j})_{j=0}^{\infty}\) such that \(N_{j}\to\infty\) and \(\min_{k}\left\{F(x_{k},T_{h}(x_{k}),b,d)\right\}\geq 0\). Now, consider the sequence \((k_{j})_{j=0}^{\infty}\) defined by \(k_{j}=\ceil{\hfrac{N_{j}}{2h}}\). As \(j\to \infty\), \(2\pi\hfrac{k_{j}}{N_{j}}\to\hfrac{\pi}{h}\). By the continuity of \(F\), \(F(x_{k_{j}},T_{h}(x_{k_{j}}),b,d)\to F(\cos\left(\hfrac{\pi}{h}\right),-1,b,d)\). Clearly, this also implies \(\limsup\left(\min_{k}\left\{F(x_{k},T_{h}(x_{k}),b,d)\right\}\right)_{j}\) is bounded above by \(F(\cos\left(\hfrac{\pi}{h}\right),-1,b,d)\). We have already established that we may choose \((b,d)\) such that \(F(\cos\left(\hfrac{\pi}{h}\right),-1,b,d)<0\) so we have a contradiction. There are some \((b,d)\) for which \(\exists M(b,d)\in \N\) with \(N\geq M(b,d)\) implying \(F(x_{k},T_{h}(x_{k}),b,d)<0\) for some \(k\), so \(N_{3}=M(b,d)\) satisfies the theorem for any such \((b,d)\). The stability of the equilibrium is entirely determined by the sign of \(F\) so this establishes the existence of \(N_{3}\) such that there is a bifurcation with respect to the continuous parameters resulting in a loss of stability.

    The bounds on which modes correspond to the largest positive eigenvalue and the first to lose stability follow from the same argument as the proof of Theorem \ref{thm: Discrete Bifurcation}. The argument in the previous proof uses properties of level sets \(F=C\) which hold for all \(C\leq0\) to establish bounds on the modes which correspond to the largest positive eigenvalue. The first mode to lose stability must have the largest eigenvalue and it must lie on the \(F=0\) level set when stability is lost, so the previous argument still applies. This establishes the bound on the first mode to lose stability and the mode corresponding to the largest positive eigenvalue and concludes the proof for the \(h\geq 3\) results.

    What remains is to prove that the \(h=1,2\) cases do not lose stability for any \((b,d)\). The \(h=1\) case is easy to show. \(T_{1}(x)=x\) so the curve \(y=T_{1}(x)\) lies entirely outside of the bounding region for \(\mathcal{R}(b,d)\) and the equilibrium is always stable. The \(h=2\) case is more difficult to show in the same direct fashion, so we instead opt for a more indirect approach. The final paragraph in the proof of Theorem \ref{thm: Asymptotes} (see appendix) establishes that the conditions for instability are equivalent to the conditions for the existence of asymptotes of the \(F=0\) level set in the \(bd\)-plane. The proof of Theorem \ref{thm: Asymptotes} does not depend on any of the facts established in the current proof so there are no concerns of inadvertently making a circular argument. The case \(h=2\) is given special consideration in the proof of Theorem \ref{thm: Asymptotes} and is shown to have no asymptotes implying no instability.
\end{proof}

\section{Discussion \& Conclusions}
We have shown that model \eqref{1D ODE} exhibits Turing instability and derived explicit formulas for the parameter regime in \(\R^{2}\times\N^{2}\) where this occurs. According to Theorem \ref{thm: Discrete Bifurcation}, choosing \((b,d)\in\mathcal{P}\), see \eqref{P Def}, guarantees that for \(h\) and \(N\) large enough, the homogeneous steady state becomes unstable to spatially periodic perturbations. Similarly, for all \(h\geq 3\) and \(N\) sufficiently large, Theorem \ref{thm: Continuous Bifurcation} states there are sets of \((b,d)\in\mathcal{P}\) where periodic patterns emerge. In addition to these results for \(N\) sufficiently large, we also established that the minimum case for pattern formation in the model is \(N=6\) with \(h=3\).

The behavior of the mean-field survival model established in this paper bears interesting similarities to lab observations. In particular, Hamada et al \cite{Hamada} observed that the long projections of melanophores were ``often greater than three times the length of the cell body and nearly half of the width of stripes". In model \eqref{1D ODE}, \(h\geq 3\) is a necessary condition for stripe formation. Additionally, the condition that \(h\) be exactly half of the width of a stripe implies that \(k=\hfrac{N}{2h}\). If \(k<\hfrac{N}{2h}\), then \(h\) will be less than half of the width of a stripe. The bounds on \(k\) in Theorems \ref{thm: Discrete Bifurcation} and \ref{thm: Continuous Bifurcation} imply that this is always the case for \(N\) sufficiently large. Additionally, empirical evidence indicates that the unstable mode with \(k\) closest to \(\hfrac{N}{2h}\) is the mode which produces the slant asymptote. This means that the regime where \(b\) is only slightly less than \(d\) is the regime that best matches this physical observation. The formula for these slant asymptotes allows this characterization to be made more precise for any chosen \(h\) or \(N\).

Mature wild-type zebrafish most commonly have four black stripes on each side of their body, indicating \(k=8\) is a particularly relevant mode to study. The linear stability analysis provides bounds, \eqref{Bounds1} and \eqref{Bounds2}, on which modes may be unstable. These bounds imply that, for \(k=8\) to be unstable, it must be that \(14<\hfrac{N}{h}\leq 32\). For \(h=3\) or \(h=14\), this corresponds to \(42< N \leq 96\) and \(196< N \leq 448\) respectively. We can compare these results to the observation in \cite{Nakamasu,Volkening2015} that the width of stripes is \(7\)-\(12\) cell diameters using the fact that a cell width should be equal to \(\hfrac{N}{2k}\). If \(k=8\) and stripes have width \(7\), then the model predicts that \(N=112\); if the width is \(12\), then \(N=192\). So, \(h=3\) is too small to match these observations and \(h=14\) is too large. So, the most realistic parameter regime for the model appears to be \(4\leq h \leq 13\), \(112\leq N \leq 192\), and \(b\) less than but approximately equal to \(d\).

These results may also be used to predict changes in pattern as a zebrafish grows. Domain growth is known to effect the development of patterns and should not be dismissed entirely. However, we may use the linear stability results as a first approximation of the true stability properties on a growing fish. The results indicate the number of stripes that are favored should grow proportional to the circumference of the fish in an approximately linear fashion. Existing data supports the idea of linear growth of stripe count relative to standard length \cite{Parichy2009}. Under the assumption that the circumference is proportional to standard length, this would also imply a linear relationship between stripe count and circumference. Without detailed data on fish circumferences or number of cells in a fish cross section, more concrete comparisons are difficult.

More generally, models which use a distant neighbor signaling interaction as a mechanism for pattern formation may follow similar trends with regards to changing the animal size. Stripe width will likely remain approximately constant while stripe count will grow. Species which demonstrate similar trends may be reasonably modelled in a similar way. Alternatively, species which are observed to have stripes which grow significantly with animal size likely cannot be modeled by this type of distant neighbor interaction or must incorporate other mechanisms.

The parameter regimes where the homogeneous steady state is unstable have explicitly defined asymptotes which bound the behavior for \((b,d)\) large. Combined with numerical calculation, this gives an effective means of approximating the linear stability properties of the system for any parameter values. We also discovered analytic bounds on the possible wavenumbers of the final pattern and may use the linear stability results as a heuristic for predicting the observed wavenumbers. 

Finally, the results for \(N\to\infty\) provide tools for approximating behavior for \(N\) large and assess the validity of continuum models. Konow et al also proposed a PDE model as a first approximation of the mean-field survival model in a continuum limit. This PDE model used a second order Taylor expansion in space to rewrite the distant neighbor interactions in terms of spatial derivatives at the point of interest \cite{Konow}. The LSA of this PDE model predicts significantly different behavior than the large \(N\) limit of the ODE model. For example, in the \(bh\)-plane, the PDE analysis predicts that the neutral stability curve will resemble a hyperbola with vertical and horizontal asymptotes and that all parameter values above the curve will be unstable \cite{Konow}. The left boundary is qualitatively similar to the ODE model, but the right bound is significantly different. The region of instability in the ODE model is bounded in the \(b\) direction by an intricate curve while the PDE model is not bounded at all. This suggests that the PDE model may only be applicable in a limited region of parameter space or that a different continuum model may more accurately approximate the ODE system. One possible explanation for this discrepancy is that a higher order Taylor series may be needed to accurately describe the distant coupling terms. It is also possible that such a Taylor series based approach is only reliable if the distance between locations goes to \(0\) as \(N\to\infty\). The author is unaware of theoretical results which delineate when such an approach may or may not be successful. Research into the conditions where such an approach produces reliable predictions may be useful for future modeling efforts. Even in contexts where one desires to study a continuum limit, performing the analysis in the discrete case can still help inform the expected properties of continuum model. 

There is still work to be done regarding the mean-field survival model. Future efforts may be directed towards analyzing the full generality of parameters in the original mean-field survival model. The effects of the other parameters may significantly change the parameter regimes in which stripes form or the wavelength of stripes which are selected. There is also potential for interesting work on the selection mechanism when multiple modes are unstable simultaneously. In numerical simulations with small amplitude and uniformly distributed noise, we see that for some parameter values only one mode is ever selected. For other parameter values, multiple modes may be selected with approximately equal likelihood depending on the initial noise. Another possible future direction is that of weakly nonlinear analysis. Near Turing bifurcations in PDEs, solutions are often well approximated by solutions where the amplitude of the unstable mode is controlled by a Ginzburg-Landau equation in appropriately rescaled variables. It is likely that a similar approach may allow for weakly nonlinear analysis near Turing bifurcations in discrete systems as well. There is also work to be done extending the model to 2D. As noted in \cite{Jewell}, non-local interactions can behave differently in different dimensions. Preliminary work on the mean-field survival model in 2D indicates that the model is also sensitive to the precise implementation of interaction with cells distance \(h\) away. Whether distance \(h\) is interpreted as only being along the coordinate axes, distance \(h\) in the 1-norm or distance \(h\) in the \(\infty\)-norm significantly changes the onset of instability. This preliminary work has also highlighted a route to a new continuum model. If \(\left\{G_{r}\right\}_{r\in\R^{+}}\) represents a family of Fourier multipliers parameterized by a characteristic interaction distance and not depending on \(b\) or \(d\), then the nearest neighbor and distant neighbor coupling terms can be replaced by \(G_{1}\) and \(G_{h}\) acting on the appropriate component for each term. The only effect this change has on the analysis in this paper is that instead of instability being controlled by the sign of \(F(x_{k},T_{h}(x_{k}),b,d)\), it is now controlled by the sign of \(F(\hat{G}_{1}(\Vec{k}),\hat{G}_{h}(\Vec{k}),b,d)\). In the analogues of Figures \ref{fig: h3N6 Cheb plots} and \ref{fig: b3d10 Cheb plots}, the curve \(y=T_{h}(x)\) would then be replaced by the implicit curve \((x,y)=(\hat{G}_{1}(\Vec{k}),\hat{G}_{h}(\Vec{k}))\). As an example, the convolution against disk or annulus shaped characteristic functions (similar to interactions in \cite{Volkening2018}) may be a reasonable continuum limit. Such an approach would also be a first step to rigorously bridging the gap between the current simplified model and the more mechanistically realistic agent based models.
\appendix
\section{Proofs of some intermediate results and Theorem \ref{thm: Asymptotes}}
\begin{proof}[Proof of Lemma \ref{lem: F Properties}]

    Recall the definition of \(F\) given below
    \begin{align*}
        F(x,y,b,d)=&-bdx^{2}+\left(bd^{2}y-b(b+d)-d(b+1)(b+2d)\right)x\\
        &+bd(b+d)y+(b+1)(b+d)(b+2d)
    \end{align*}
    
    We will first show that the set \(\mathcal{R}(b,d)=\left\{(x,y)\in [-1,1]^{2}:F(x,y,b,d)<0 \right\}\) is empty for \((b,d)\notin \mathcal{P}\) and contained in the prescribed region for \((b,d)\in\mathcal{P}\). The set \(\mathcal{R}(b,d)\) is the complement in \([-1,1]^{2}\) of \(\bigcup_{y\in[-1,1]}\left\{x\in\R:F(x,y,b,d)\geq0\right\}\), so establishing properties of the sets \(\left\{x\in\R:F(x,y,b,d)\geq0\right\}\) will allow us to establish properties of \(\mathcal{R}(b,d)\). To do this, we will need the following derivatives:
    \begin{align*}
        \frac{\partial F(x,y,b,d)}{\partial y}=&bd(dx+b+d).\\
         \frac{\partial^{2}F(x,y,b,d)}{\partial x^{2}}=&-2bd.
    \end{align*}
    For the admitted values of \((b,d)\) and \(x\), \(0<b^{2}d\leq \frac{\partial F(x,y,b,d)}{\partial y}\) and \(\frac{\partial^{2}F(x,y,b,d)}{\partial x^{2}}<0\). The concavity in \(x\) means the set \(\left\{x\in\R:F(x,y,b,d)\geq0\right\}\) is a single closed interval for all fixed \(y\). We now compute \(F(-1,-1,b,d)\), \(F(0,-1,b,d)\) and \(F(1,-1,b,d)\) to establish bounds on these intervals.
    \begin{align*}
        F(-1,-1,b,d)=&b^{3}+3b^{2}d+4bd^{2}+2b^{2}+4bd+4d^{2}.\\
        F(0,-1,b,d)=&b(b+d)^{2}+(b+d)(b+2d).\\
        F(1,-1,b,d)=&b(b+2d)(b-d).
    \end{align*}
    
    We see that \(F(0,-1,b,d)>0\) and \(F(1,-1,b,d)<0\) for all \((b,d)\in\mathcal{P}\). This shows that if \((b,d)\in\mathcal{P}\), there is a root of \(F(x.-1,b,d)\) with \(0<x<1\). We also see that \(F(-1,-1,b,d)>0\). Because \(F\) is strictly increasing in \(y\) the intervals \(\left\{x\in\R:F(x,y,b,d)\geq0\right\}\) contain subintervals of the form \([-1,R]\). This is important to note because it establishes that \(\mathcal{R}(b,d)\) must be a single contiguous region on the right half of the box \([-1,1]^{2}\). Additionally, the fact that \(F\) is strictly increasing in \(y\) guarantees the leftmost point in \(\mathcal{R}(b,d)\) has \(y=-1\). This means that a lower bound in \(x\) on the root of \(F(x,-1,b,d)\) also provides a lower bound in \(x\) of \(\mathcal{R}(b,d)\). We obtain the this bound by computing the root of the secant line connecting \(F(0,-1,b,d)\) and \(F(1,-1,b,d)\). The concavity of \(F\) in \(x\) guarantees this is a lower bound on the actual root. The lower bound is given by
    \begin{align*}
        \frac{F(0,-1,b,d)}{F(0,-1,b,d)-F(1,-1,b,d)}=&\frac{b(b+d)^{2}+(b+d)(b+2d)}{bd(b+3d)+(b+d)(b+2d)}.
    \end{align*}
    An alternative characterization is
    \begin{align*}
        \frac{F(0,-1,b,d)}{F(0,-1,b,d)-F(1,-1,b,d)}=&1+\frac{F(1,-1,b,d)}{F(0,-1,b,d)-F(1,-1,b,d)},\\
        =&1-\frac{b(b+2d)(d-b)}{b^{2}d+3bd^{2}+b^{2}+3bd+2d^{2}}.
    \end{align*}
    From these two equivalent presentations, it is clear that the bound lies between \(0\) and \(1\) as expected. This establishes the \(x\) bounds on the region \(\mathcal{R}(b,d)\).
    
    We establish the \(y\) bounds on \(\mathcal{R}(b,d)\) by characterizing when the closed intervals \(\left\{x\in\R:F(x,y,b,d)\geq0\right\}\)  contain all of \([-1,1]\). We have already established that these intervals contain subintervals of the form \([-1,R]\) so they contain all of \([-1,1]\) if and only if \(F(1,y,b,d)\geq 0\). Since \(F\) is increasing in \(y\), the root in \(y\) of \(F(1,y,b,d)\) is an upper bound on \(\mathcal{R}(b,d)\). The formula for \(F(1,y,b,d)\) is
    \begin{align*}
        F(1,y,b,d)=&b(b+2d)(dy+b),
    \end{align*}
    so \(y=\frac{-b}{d}\) is an upper bound on \(\mathcal{R}(b,d)\). This establishes the desired bounds on \(\mathcal{R}(b,d)\) in the case that \((b,d)\in\mathcal{P}\). If \((b,d)\notin\mathcal{P}\), then the upper bound on \(\mathcal{R}(b,d)\) is below the box \([-1,1]^{2}\) so \(\mathcal{R}(b,d)\) is empty. This establishes the first and fourth bullet points in the lemma.

    To show the second bullet point is true, we must establish an interval \((L,1]\) which is contained in \(\left\{x\in\R:F(x,-1,b,d)<0\right\}\) and find the infimum of such \(L\) as \((b,d)\) range over \(\mathcal{P}\). We do this by computing the tangent line to \(F(x,-1,b,d)\) at \(x=1\). Because \(F\) is concave down in \(x\), the root of this tangent line will be a strict upper bound on the root of \(F(x,-1,b,d)\). The tangent line is
    \begin{align*}
        y=&\left(-b^{2}d-3bd^{2}-b^{2}-4bd-2d^{2}\right)(x-1)+b^{3}+b^{2}d-2bd^{2}.
    \end{align*}
    The root of this tangent line is 
    \begin{align*}
        L=&\frac{b^{3}+2b^{2}d+bd^{2}+b^{2}+4bd+2d^{2}}{b^{2}d+3bd^{2}+b^{2}+4bd+2d^{2}}.
    \end{align*}
    The analysis for bullet point one shows that \(F(x,-1,b,d)=0\) defines the leftmost edge of \(\left\{x\in\R:F(x,-1,b,d)<0\right\}\) so \((L,1]\) is contained in \(\left\{x\in[-1,1]:F(x,-1,b,d)<0\right\}\) as desired.

    To establish the infimum of \(L\), we begin by defining \(\alpha=\hfrac{b}{d}\) and rewriting \(L\) accordingly.
    \begin{align*}
        L=&\frac{\alpha(\alpha+1)^{2}d+\alpha^{2}+4\alpha+2}{\alpha(\alpha+3)d+\alpha^{2}+4\alpha+2}.
    \end{align*}
    The set \(\mathcal{P}\) corresponds to the set \(d>0\) and \(0<\alpha <1\). To determine the infimum over this set of \(\alpha\) and \(d\) values, we need the partial derivatives of \(L\).
    \begin{align*}
        \frac{\partial L(\alpha,d)}{\partial d}=&\frac{\alpha(\alpha^{2}+4\alpha+2)(\alpha^{2}+\alpha-1)}{\left(\alpha(\alpha+3)d+\alpha^{2}+4\alpha+2\right)^{2}}.\\
        \frac{\partial L(\alpha,d)}{\partial \alpha}=&\frac{d\left(\alpha^{2}(\alpha^{2}+8\alpha+5)d+\alpha^{4}+8\alpha^{3}+12\alpha^{2}-8\alpha+8\right)}{\left(\alpha(\alpha+3)d+\alpha^{2}+4\alpha+2\right)^{2}}.
    \end{align*}
    
    We now study the signs of each partial derivative. As the denominators are always positive, we only study the numerators. For \(\tfrac{\partial L}{\partial d}\), we simply factor and check signs.
    \begin{align*}
        \alpha(\alpha^{2}+4\alpha+2)(\alpha^{2}+\alpha-1)=&\alpha(\alpha+2+\sqrt{2})(\alpha+2-\sqrt{2})\cdot\\
        &\cdot(\alpha+\tfrac{1-\sqrt{5}}{2})(\alpha+\tfrac{1+\sqrt{5}}{2}).
    \end{align*}
    So, \(\tfrac{\partial L}{\partial d}<0\) for \(0<\alpha<\tfrac{-1+\sqrt{5}}{2}\) and \(\tfrac{\partial L}{\partial d}>0\) for \(\tfrac{-1+\sqrt{5}}{2}<\alpha\leq 1\).

    For \(\tfrac{\partial L}{\partial \alpha}\), we first note that the numerator is a positive quadratic with respect to \(d\) so it is positive outside of the roots and negative between the roots. Additionally, one of the roots is exactly at \(d=0\). So, knowing the other root is sufficient for determining the sign of \(\tfrac{\partial L}{\partial \alpha}\). The other root is
    \begin{align*}
        d=&-\frac{\alpha^{4}+8\alpha^{3}+12\alpha^{2}-8\alpha+8}{\alpha^{2}(\alpha^{2}+8\alpha+5)}
    \end{align*}
    The numerator is greater than or equal to \(8(1-\alpha)\). This means that the numerator is non-negative and the root is non-positive. So, \(\tfrac{\partial L}{\partial \alpha}>0\) in the relevant regime.

    This analysis of partial derivatives indicates that to find the infimum of \(L\), we must study a limit with \(\alpha\to 0\). As alpha decreases, we eventually enter the regime where \(\tfrac{\partial L}{\partial d}<0\). This means we must also consider a limit as \(d\to \infty\) for the infimum. This produces a limit, 
    \begin{align*}
        \lim_{(\alpha,d)\to(0,\infty)}\frac{\alpha(\alpha+1)^{2}d+\alpha^{2}+4\alpha+2}{\alpha(\alpha+3)d+\alpha^{2}+4\alpha+2},
    \end{align*}
    whose value is not well defined. To resolve this issue, we consider the family of sublimits defined by the constraints
    \begin{align*}
        \lim_{t\to\infty}(\alpha(t),d(t))=(0,\infty),\\
        (\alpha',d')=\frac{-\nabla L}{\abs{\abs{L}}}.
    \end{align*}
    This is well-defined because on the relevant domain \(\nabla L\) is non-vanishing. These constraints mean that the sublimits follow the gradient backwards towards the limit point and are parameterized by arc length. If we consider an arbitrary sublimit, \((\alpha_{0}(t),d_{0}(t))\), parameterized by arc length but which is not in this family, then we may choose an element of our family, \((\alpha_{1}(t),d_{1}(t))\) by specifying a parameter value, \(T\) such that \((\alpha_{0}(T),d_{0}(T))=(\alpha_{1}(T),d_{1}(T))\). Clearly, \(\forall t\geq T\), \(L(\alpha_{0}(t),d_{0}(t))\geq L(\alpha_{1}(T),d_{1}(T))\). Thus, \(\lim_{t\to\infty}L(\alpha(t),d(t))\) is minimized by some member of our family. We may use L'Hopital's rule to evaluate any sublimit in this family. If we define \(g(\alpha,d)=\alpha(\alpha+1)^{2}d+\alpha^{2}+4\alpha+2\) ,the numerator of our limit, and \(h(\alpha,d)=\alpha(\alpha+3)d+\alpha^{2}+4\alpha+2\), the denominator of our limit, then we compute
    \begin{align*}
        \lim_{t\to\infty}L(\alpha(t),d(t))=&\lim_{t\to\infty}\frac{\left(\nabla g\cdot \hfrac{-\nabla L}{\abs{\abs{L}}}\right)}{\left(\nabla h\cdot \hfrac{-\nabla L}{\abs{\abs{L}}}\right)},
    \end{align*}
    if the limit exists. This may be rewritten as
    \begin{align*}
        \lim_{t\to\infty}L(\alpha(t),d(t))=&\frac{\left(\lim_{t\to\infty}\tfrac{\hfrac{\partial g}{\partial \alpha}}{\hfrac{\partial h}{\partial \alpha}}\right)-\left(\lim_{t\to\infty}\tfrac{\hfrac{\partial g}{\partial d}}{\hfrac{\partial h}{\partial \alpha}}\right)\left(\lim_{t\to\infty}\tfrac{\hfrac{\partial L}{\partial d}}{\hfrac{\partial L}{\partial \alpha}}\right)}{1-\left(\lim_{t\to\infty}\tfrac{\hfrac{\partial h}{\partial d}}{\hfrac{\partial h}{\partial \alpha}}\right)\left(\lim_{t\to\infty}\tfrac{\hfrac{\partial L}{\partial d}}{\hfrac{\partial L}{\partial \alpha}}\right)},
    \end{align*}
    once again assuming that all limits exist. We compute the relevant fractions for this limit as
    \begin{align*}
        \frac{\hfrac{\partial g}{\partial \alpha}}{\hfrac{\partial h}{\partial \alpha}}=&\frac{(\alpha+1)(3\alpha+1)d+2\alpha+4}{(2\alpha+3)d+2\alpha+4},\\
        \frac{\hfrac{\partial g}{\partial d}}{\hfrac{\partial h}{\partial \alpha}}=&\frac{\alpha(\alpha+1)^{2}}{(2\alpha+3)d+2\alpha+4},\\
        \frac{\hfrac{\partial L}{\partial d}}{\hfrac{\partial L}{\partial \alpha}}=&\frac{\alpha(\alpha^{2}+4\alpha+2)(\alpha^{2}+\alpha-1)}{d\left(\alpha^{2}(\alpha^{2}+8\alpha+5)d+\alpha^{4}+8\alpha^{3}+12\alpha^{2}-8\alpha+8\right)},\\
        \frac{\hfrac{\partial h}{\partial d}}{\hfrac{\partial h}{\partial \alpha}}=&\frac{\alpha(\alpha+3)}{(2\alpha+3)d+2\alpha+4}.
    \end{align*}
    Recalling that as \(t\to \infty\), \(\alpha(t)\to 0\) and \(d(t)\to \infty\), we are now able to verify all relevant limits exist and compute the infimum of \(L\).
    \begin{align*}
        \lim_{t\to\infty}L(\alpha(t),d(t))=&\frac{\hfrac{1}{3}-0\cdot 0}{1-0\cdot 0}=\hfrac{1}{3}.
    \end{align*}
    This establishes the infimum of \(L\) and proves the second bullet point of the lemma as desired.
    
    Finally, we prove the third bullet point in the lemma. The level sets \(F(x,y,b,d)=C\) can be described explicitly by the formula
    \begin{align*}
        y=&\tfrac{1}{bd}\left(bx+(b+1)(b+2d)-\frac{2(b+1)(b+d)(b+2d)-C}{dx+b+d}\right).
    \end{align*}
    For \((b,d)\in\mathcal{P}\) and \(x\in[0,1]\), this function is easily verified to be concave down and increasing whenever \(C\leq 0\). Because the region \(\mathcal{R}(b,d)\) is exactly the set of points in \([-1,1]^{2}\) with \(C<0\), the region must be convex. This establishes the properties of the level sets of \(F\) in the \(xy\)-plane and completes the proof.
\end{proof}

The proof of Theorem \ref{thm: Asymptotes} makes use of the following lemma about the formulation of asymptotes of curves expressed in polar coordinates.

\begin{lemma}\label{lem: Polar Asym}
    Let \(r(\theta)\) define a curve in polar coordinates and let \(\theta_{*}\) be an angle such that \(\lim_{\theta\to \theta_{*}}\abs{r(\theta)}=\infty\). \(r(\theta)\) has a straight-line asymptote parallel to the line \(\tan(\theta)=\tan(\theta_{*})\) if \(r(\theta)\) admits and asymptotic expansion near \(\theta_{*}\) of the form
    \begin{align*}
        r(\theta_{*}+\varepsilon)=\tfrac{\alpha}{\varepsilon}+\beta+\O(\varepsilon)
    \end{align*}
\end{lemma}

This lemma will later allow us to verify that the implicitly defined curve \(F(x,T_{h}(x),r\cos(\theta),r\sin(\theta))\) has straight-line asymptotes for the values of \(\theta\) where \(r\to\infty\).

\begin{proof}[Proof of Lemma \ref{lem: Polar Asym}]
    Assume \(r(\theta)\) admits an asymptotic expansion of the form
    \begin{align*}
        r(\theta_{*}+\varepsilon)=&\tfrac{\alpha}{\varepsilon}+\beta+\O(\varepsilon).
    \end{align*}
    We can directly compute a parametric description of the Cartesian coordinates of the curve using appropriate Taylor series.
    \begin{align*}
        x=&\frac{\alpha\cos(\theta_{*})}{\varepsilon}-\alpha\sin(\theta_{*})+\beta\cos(\theta_{*})+\O(\varepsilon).\\
        y=&\frac{\alpha\sin(\theta_{*})}{\varepsilon}+\alpha\cos(\theta_{*})+\beta\sin(\theta_{*})+\O(\varepsilon).
    \end{align*}
    This can be rewritten in slope-intercept form as a function of only \(\alpha\) for \(\theta_{*}\neq \hfrac{\pi}{2}+n\pi\).
    \begin{align*}
        y=&\tan(\theta_{*})x+\alpha\sec(\theta_{*})
    \end{align*}
    For \(\theta_{*}=\hfrac{\pi}{2}+n\pi\), the line is \(x=(-1)^{n}\alpha\). This completes the proof.
\end{proof}

\begin{proof}[Proof of Theorem \ref{thm: Asymptotes}]

    Recall that linear stability in the \(k\) Fourier mode is controlled by the sign of \(F(x_{k},T_{h}(x_{k}),b,d)\) and that the alternative formula for \(F\), given in equation \ref{F Alt}, states
    \begin{align*}
        F(x_{k},T_{h}(x_{k}),b,d)=&b\left(b^{2}+f_{1}(x_{k})bd+f_{2}(x_{k})d^{2}\right)\\
        &-(x_{k}-1)\left(b^{2}+(x_{k}+3)bd+2d^{2}\right),
    \end{align*}
    where \(f_{1}(x)=-x+T_{h}(x)+3\) and \(f_{2}(x)=(x+1)T_{h}(x)-2(x-1)\).
    
    We are interested in the asymptotes of \(F(x_{k},T_{h}(x_{k}),b,d)=0\) in the \(bd\)-plane for the values of \(k\) for which the equilibrium is unstable. We first study the asymptotes of \(F(x,T_{h}(x),b,d)=0\), then verify that the conditions on \(x\) which produce asymptotes are the same conditions which produce instability, and finally restrict the value of \(x\) to those \(x_{k}\) of interest. The significance of this second step is that we have no a priori knowledge that the set of \((b,d)\) values which have \(F(x,T_{h}(x),b,d)<0\) is not bounded or that the curve \(F(x,T_{h}(x),b,d)=0\) does not asymptotically resemble a parabola or other more complicated curve.
    
    Defining \(b=r\cos(\theta)\) and \(d=r\sin(\theta)\), we find that
    \begin{align*}
        F(x,T_{h}(x),r\cos(\theta),r\sin(\theta))=&r^{2}\left(r\cos(\theta)\left(\cos^{2}(\theta)+f_{1}(x)\cos(\theta)\sin(\theta)\right.\right.\\
        &\left.\left.+f_{2}(x)\sin^{2}(\theta)\right)-(x-1)\left(\cos^{2}(\theta)\right.\right.\\
        &\left.\left.+(x+3)\cos(\theta)\sin(\theta)+2\sin^{2}(\theta)\right)\right).
    \end{align*}
    For \((b,d)\) to be admissible, we require either \(0<\theta<\hfrac{\pi}{2}\) and \(r>0\) or \(\pi<\theta<\hfrac{3\pi}{2}\) and \(r<0\). Thus, the portion of interest of the implicit curve \(F=0\) may be given explicitly by
    \begin{align*}
        r(\theta)=&\frac{(x-1)Q_{1}(\cot(\theta))}{\cos(\theta)Q_{2}(\cot(\theta))}.\\
        Q_{1}(z)=&z^{2}+(x+3)z+2.\\
        Q_{2}(z)=&z^{2}+f_{1}(x)z+f_{2}(x).
    \end{align*}
    Note that the conditions \(0<\theta<\hfrac{\pi}{2}\) and \(r>0\) and conditions \(\pi<\theta<\hfrac{3\pi}{2}\) and \(r<0\) define the same points in the plane as the sign of each \(\cos(\theta)\) and \(\sin(\theta)\) term change between the different ranges of \(\theta\). As such, we will assume \(0<\theta<\hfrac{\pi}{2}\) in the remaining analysis. We still require that \(r>0\) and will address this question as part of our analysis of the angles where asymptotes exist.

    To find the values of \(\theta\) for which \(r\to\infty\), we seek values where the denominator equal to \(0\) but the numerator is not. Clearly, \(\theta=\hfrac{\pi}{2}\) is one such angle. We must also study the roots and signs of \(Q_{1}(\cot(\theta))\) and \(Q_{2}(\cot(\theta))\) to determine if there are other angles where \(\abs{r}\to \infty\) and to determine when \(r\) is positive.

    The derivative of \(Q_{1}(z)\) is
    \begin{align*}
        \tfrac{dQ_{1}}{dz}=&2z+x+3.
    \end{align*}
    Our interest is only on the domain \(x\in[-1,1]\) and \(\cot(\theta)\geq0\) so the numerator of \(r\) is increasing with respect to \(\cot(\theta)\). When \(\cot(\theta)=0\), \(Q_{1}(\cot(\theta))\) is positive, so it is positive for the entirety of our domain of interest. So, any angle where \(Q_{2}(\cot(\theta))=0\) has \(\abs{r}\to\infty\). Additionally, within the region \(0<\theta<\tfrac{\pi}{2}\), the angles where \(r\) is positive are exactly those where \(Q_{2}(\cot(\theta))<0\). This follows from the observation that \(x-1\leq0\) and \(\cos(\theta)>0\) for the values of interest.
    
    To determine when \(Q_{2}(\cot(\theta))=0\), we will study the signs of \(f_{1}(x)\) and \(f_{2}(x)\). \(f_{1}(x)<0\) if and only if \(T_{h}(x)>x-3\). However, \(T_{h}(x)\geq -1\) and \(x-3\leq -2\) for \(x\in[-1,1]\). The linear coefficient of \(Q_{2}\) is positive, so there will be exactly one positive root if and only if \(f_{2}(x)<0\).

    \(f_{2}(x)<0\) if and only if \(T_{h}(x)<2\tfrac{x-1}{x+1}\). The function \(2\tfrac{x-1}{x+1}\) is increasing on the relevant domain and is equal to \(-1\) at \(x=\hfrac{1}{3}\). Since \(T_{h}(x)\) have a minimum value of \(-1\) on our domain, we find that \(x<\hfrac{1}{3}\) implies that \(f_{2}(x)\) is positive. Additionally, \(T_{h}(x)\) having a local minimum with \(x\) above \(\hfrac{1}{3}\) is sufficient to guarantee an interval of \(x\) values for which \(f_{2}(x)<0\). This occurs exactly when \(h\geq 3\). So, for all \(h\geq 3\), there is a subinterval of \(x\)-values in \((\hfrac{1}{3},1]\) where \(Q_{2}(\cot(\theta))=0\) admits positive solutions. The \(h=1,2\) cases can be checked individually. 
    
    We begin with the \(h=1\) case. \(T_{1}(x)=x\) so we compute
    \begin{align*}
        f_{2}(x)=&x^{2}-x+2.
    \end{align*}
    This can be factored as \(f_{2}(x)=(x-\tfrac{1+\sqrt{7}i}{2})(x-\tfrac{1-\sqrt{7}i}{2})\) so, the case \(h=1\) does not allow \(Q_{2}(z)\) to have any positive roots.

    For the \(h=2\) case, we use \(T_{2}(x)=2x^{2}-1\) to compute
    \begin{align*}
        f_{2}(x)=&2x^{3}+2x^{2}-3x+1.
    \end{align*}
    To determine whether or not this cubic has a positive, real root, we note that the cubic coefficient is positive and the constant term is positive. As such, the only way to have a positive real root is for the local minimum value to be less than or equal to \(0\). Denoting this cubic as \(P(x)=2x^{3}+2x^{2}-3x+1\), we compute \(P'(x)=6x^{2}+4x-3\). The roots of \(P'\) are \(x=\frac{-1\pm\sqrt{\hfrac{11}{2}}}{3}\) so the local minimum occurs at \(x=\frac{-1+\sqrt{\hfrac{11}{2}}}{3}\). Computing the value, we find
    \begin{align*}
        P\left(\frac{-1+\sqrt{\hfrac{11}{2}}}{3}\right)=&\tfrac{1}{27}\left(58-22\sqrt{\hfrac{11}{2}}\right)>0.
    \end{align*}
    Thus, the \(h=2\) case also disallows \(Q_{2}(z)\) from having positive roots.
    
    So, we conclude that \(h\geq 3\) is necessary and sufficient for the admission of positive roots of \(Q_{2}(z)\). The coefficient of \(z^{2}\) in \(Q_{2}\) is positive and the analysis of \(f_{1}(x)\) and \(f_{2}(x)\) show that there is one root which is positive and one root which is negative, so \(Q_{2}(z)<0\) on the interval \([0,z_{1})\) where \(z_{1}\) is the positive root. We have previously established that for all \(0<\theta<\tfrac{\pi}{2}\), \(r>0\) if \(Q_{2}(\cot(\theta))<0\), so \(\cot(\theta)=z_{1}\) establishes a lower bound on \(\theta\). The angles for which \(r>0\) are \(\theta_{1}<\theta<\hfrac{\pi}{2}\) where \(\theta_{1}\) is characterized using the quadratic formula.
    \begin{align*}
        2\cot(\theta_{1})=&-T_{h}(x)+x-3+\sqrt{T_{h}^{2}(x)-6xT_{h}(x)+2T_{h}(x)+(x+1)^{2}}.
    \end{align*}

    To summarize the results so far in the proof, we now know that the curve \(F(x,T_{h}(x),r\cos(\theta),r\sin(\theta))=0\) can be expressed with \(r\) as a function of \(\theta\). If \(h\geq 3\), there is a subinterval of \(x\) values with the following properties: The curve \(F=0\) exists in the first quadrant and there are two candidate angles, \(\theta=\hfrac{\pi}{2},\theta_{1}\), which may correspond to straight-line asymptotes. This subinterval is always contained in \((\hfrac{1}{3},1]\). If \(h<3\), then there is no such subinterval.

    We now use Lemma \ref{lem: Polar Asym} to show that these candidate angles produce actual asymptotes. We prepare by rewriting the expression for \(r\) as
    \begin{align*}
        r(\theta)=&\frac{(x-1)\left((x+3)\sin(2\theta)-\cos(2\theta)+3\right)}{\cos(\theta)\left(f_{1}(x)\sin(2\theta)-(f_{2}(1)-1)\cos(2\theta)+f_{2}(x)+1\right)}.
    \end{align*}
    Using Taylor expansions of \(\cos(\theta)\), \(\cos(2\theta)\), and \(\sin(2\theta)\) around arbitrary \(\theta_{*}\), we find
    \begin{align*}
        r(\theta_{*}+\varepsilon)=&\frac{(x-1)\left((x+3)\sin(2\theta_{*})-\cos(2\theta_{*})+3+\O(\varepsilon)\right)}{j(x,\theta_{*},\varepsilon)}.\\
        j(x,\theta_{*},\varepsilon)=&\cos(\theta_{*})\left(f_{1}(x)\sin(2\theta_{*})-(f_{2}(1)-1)\cos(2\theta_{*})+f_{2}(x)+1\right)\\
        &+\Big(\sin(\theta_{*})\big(f_{1}(x)\sin(2\theta_{*})-(f_{2}(1)-1)\cos(2\theta_{*})+f_{2}(x)+1\big)\\
        &+2\cos(\theta_{*})\big(f_{1}(x)\cos(2\theta_{*})+(f_{2}-1)\sin(2\theta_{*})\big)\Big)\varepsilon+\O(\varepsilon^{2}).
    \end{align*}

    For the case \(\theta_{*}=\hfrac{\pi}{2}\), we find \(r\) admits an asymptotic expansion of the form 
    \begin{align*}
        r\left(\hfrac{\pi}{2}+\varepsilon\right)=&\frac{-2(x-1)}{f_{2}(x)\varepsilon}+\O(1).
    \end{align*}
    For the case \(\theta_{*}=\theta_{1}\), we find \(r\) admits an asymptotic expansion of the form
    \begin{align*}
        r(\theta_{1}+\varepsilon)=&\frac{(x-1)\left((x+3)\sin(2\theta_{1})-\cos(2\theta_{1})+3\right)}{2\cos(\theta_{1})\left(f_{1}(x)\cos(2\theta_{1})+(f_{2}(x)-1)\sin(2\theta_{1})\right)\varepsilon}+\O(1).
    \end{align*}

    So, both \(\theta=\hfrac{\pi}{2}\) and \(\theta=\theta_{1}\) correspond to straight-line asymptotes. The proof of Lemma \ref{lem: Polar Asym} includes an explicit formula for the asymptotes in Cartesian coordinates using the coefficient of \(\hfrac{1}{\varepsilon}\) in the asymptotic expansion. This concludes the characterization of the straight-line asymptotes of \(F(x,T_{h}(x),b,d)=0\).

    We now turn our focus to showing these conditions are the same as the equilibrium being unstable. The previous analysis shows that the curve \(F(x,T_{h}(x),b,d)=0\) in the \(bd\)-plane always has straight-line asymptotes when it exists in the first quadrant. Since this curve is the boundary of the set \(\left\{(b,d):F(x,T_{h}(x),b,d)<0\right\}\), either the entire first quadrant has \(F<0\), or the set having elements in the first quadrant is equivalent to the existence of the asymptotes. We know from the proof of Theorem \ref{thm: Discrete Bifurcation}, that this first option is impossible, because the set \(\left\{(b,d):F(x,T_{h}(x),b,d)<0\right\}\) is a subset of \(\mathcal{P}\). For the same reason, we know that the set lies right of the vertical asymptote and above the slant asymptote, or else it would extend beyond \(\mathcal{P}\). This may also be shown directly by computing \(\tfrac{\partial F}{\partial r}\) and noting that our previous analysis shows \(F=0\) implies \(\tfrac{\partial F}{\partial r}<0\) and the set must be further from the origin than the curve. This establishes all of the desired claims and concludes the proof.
\end{proof}

The final paragraph shows that the proof of Theorem \ref{thm: Asymptotes} can also function as an indirect method to prove Theorems \ref{thm: Discrete Bifurcation} and \ref{thm: Continuous Bifurcation}, but without the visual intuition for the effects of changing parameters. This is worth noting as there may be other similar models for which one approach generalizes more easily than the other.

Corollary \ref{cor: Asymptotes} follows immediately from Theorem \ref{thm: Asymptotes} by taking the most extreme cases for each asymptote type. The left-most vertical asymptote will have the minimum value. The slant asymptote which is eventually the right-most bound will be the one with the smallest slope and therefore the maximum value of \(\cot(\theta_{k})\).

Corollary \ref{cor: Large N Bifurcation} is a consequence of the previous results and continuity. Only the \(k\) satisfying the bounds in inequality \eqref{Bounds2} can have non-empty \(\mathcal{U}_{k}(h,N)\), so the union over \(I\) clearly covers the union over relevant \(k\) if \(I\) is chosen appropriately. As \(N\to\infty\), the values \(\hfrac{k}{N}\) become dense in \([0,1]\) in the sense that any fixed interval will contain \(\frac{k}{N}\) for some \(k\) and \(N\) sufficiently large. This means that the unions are equivalent in the limit for appropriate \(I\). The density of \(\hfrac{k}{N}\) implies that \(x_{k}\) become dense in \([-1,1]\). The alternative characterization of the limiting bifurcation curve comes from imposing the condition that \(F(x,T_{h}(x),b,d)=0\) for some \(x\in[-1,1]\) but \(F(z,T_{h}(z),b,d)\neq 0\) for \(z\neq x\) and the observation that \(F(x,T_{h}(x),b,d)\) is differentiable in \(x\). Finally, the asymptotes follow from the same density argument as above applied to the asymptote formulae.

\section*{Acknowledgements}
The author would like to thank Tasso Kaper, Irv Epstein, Richard Bertram, and Sam Isaacson for their helpful feedback on the presentation of this paper. The author would also like to thank the anonymous reviewers of this paper for their helpful feedback.

\section*{Funding}
This work was partially funded by the grant NSF DMS 1616064.

\bibliography{Zebrafish.bib}
\bibliographystyle{elsarticle-num.bst}

\end{document}